\documentclass[a4paper,11pt]{article}
\pdfoutput=1
\usepackage{hyperref}
\usepackage{epsfig}
\usepackage{amsmath}
\usepackage{amssymb}
\usepackage{color}
\usepackage[footnotesize]{caption}
\usepackage[makeroom]{cancel}
\usepackage{cite}
\usepackage{graphicx}
\usepackage{multirow}
\usepackage{feynmf}
\usepackage{pdfpages}

\newcommand{\eVq}  {\text{eV}^2}
\newcommand{\mb}{\mbox{\tiny$\Delta$}}
\newcommand{\R}{\mbox{\tiny$R$}}
\newcommand{\A}{\mbox{\tiny$A$}}
\newcommand{\B}{\mbox{\tiny$B$}}
\newcommand{\D}{\mbox{\tiny$D$}}
\newcommand{\T}{\mbox{\tiny$T$}}
\newcommand{\sm}{\mbox{\tiny$S$}}

\newcommand{\xx}{\mbox{\tiny$\xi$}}
\newcommand{\x}{\mbox{\tiny$\xi'$}}
\newcommand{\hh}{\mbox{\tiny$H$}}
\newcommand{\ii}{\mbox{\tiny$I$}}
\newcommand{\iii}{\mbox{\tiny$II$}}
\newcommand{\N}{\mbox{\tiny$N$}}
\newcommand{\Ll}{\mbox{\tiny$L$}}

\numberwithin{equation}{section}

\begin{document}

\begin{center}
{\Large\bf Spontaneous CP Violation in Lepton-sector: a common origin for 
$\theta_{13}$, Dirac CP phase and leptogenesis}
\\[2mm]
\vskip 2cm

{ Biswajit Karmakar$^{a,}$\footnote{k.biswajit@iitg.ernet.in},
Arunansu Sil$^{a,}$\footnote{asil@iitg.ernet.in}}\\[3mm]
{\it{
$^a$ Indian Institute of Technology Guwahati, 781039 Assam, India}
}
\end{center}

\vskip 1cm

\begin{abstract}

\noindent A possible interplay between the two terms of the general 
type-II seesaw formula is exercised which leads to the generation of nonzero 
$\theta_{13}$. The specific flavor structure of the model, guided 
by the $A_4 \times Z_4 \times Z_3$ symmetry and accompanied 
with the Standard Model singlet flavons, yields the conventional 
seesaw contribution to produce the tribimaximal lepton mixing which 
is further corrected by the presence of the $SU(2)_L$ triplet contribution 
to accommodate $\theta_{13}$. We consider the CP symmetry to be 
spontaneously broken by the complex vacuum expectation value (vev) 
of a singlet field $S$. While the magnitude of its complex vev is responsible 
for generating $\theta_{13}$, 
its phase part induces the low energy CP violating phase ($\delta$) 
and the CP violation required for leptogenesis. Hence the triplet 
contribution, although sub-dominant, plays crucial role in providing 
a common source for non-zero $\theta_{13}$, $\delta$ and CP-violation 
required for leptogenesis. We find that the recent hint for $\delta$ close to $3\pi/2$ is
somewhat favored in this set-up
though it excludes the exact equality with $3\pi/2$. We also discuss the generation of 
lepton asymmetry in this scenario. 

\end{abstract}


\section{Introduction}\label{sec1}
 The question whether there exists an underlying principle to
understand the pattern of lepton mixing, which is quite 
different from the quark mixing, demands the study of neutrino
mass matrix as well as the charged lepton one into a deeper 
level. The smallness of neutrino masses can be well understood
by the seesaw mechanism in a natural way. Type-I seesaw mechanism 
\cite{Minkowski:1977sc,GellMann:1980vs,Mohapatra:1979ia,
Yanagida:1980xy}
provides the simplest possibility by extending the
Standard Model (SM) with three right-handed (RH) neutrinos.
An introduction of discrete symmetries into it may reveal
the flavor structure of the neutrino and charged lepton mass matrix.
For example, a type-I seesaw in conjugation with $A_4$ explains the 
tribimaximal lepton mixing pattern (TBM) \cite{Harrison:1999cf}
in presence of SM singlet flavon (charged under $A_4$) fields 
which get vacuum expectation values (vev)\cite{Ma:2004zv,
Altarelli:2005yp,Altarelli:2005yx}. However the original
approach fails to accommodate the recent observation of
non-zero $\theta_{13}$\cite{Abe:2011fz,DayaBay,Ahn:2012nd,Abe:2013hdq}.
In \cite{Karmakar:2014dva}, we have shown that an extension of the
Altarelli-Feruglio (AF) model \cite{Altarelli:2005yx} by one additional    
flavon field can be employed to have a nonzero $\theta_{13}$ 
consistent with the present experimental results. The set-up 
also constraints the two Majorana phases involved in the lepton 
mixing matrix. The deviation of the TBM pattern is achieved through a
deformation of the RH neutrino mass matrix compared to the
original one. 
On the other hand, within the framework of a general type-II seesaw 
mechanism (where both RH neutrinos and $SU(2)_L$ triplet Higgs are present), 
light neutrino mass depends upon comparative magnitude of the pure 
type-I (mediated by heavy RH neutrinos) and triplet contributions. 
This interplay is well studied in the literature \cite{Joshipura:1999is,
Joshipura:2001ya,Bajc:2002iw, Antusch:2004xd,Antusch:2004xy,Sahu:2004ny,
Rodejohann:2004qh,Chen:2005jm, Bertolini:2005qb,Akhmedov:2006de}.
In recent years keeping in mind that $\theta_{13}$ is nonzero,
efforts have been given to realize leptogenesis \cite{Rodejohann:2004cg,
Gu:2006wj,Abada:2008gs,Borah:2013bza,Borah:2014fga,Borah:2014bda,Kalita:2014vxa} and 
linking it with $\theta_{13}$ in models based on type-II seesaw\cite{Pramanick:2015qga}.

In this article, we focus on the generation of light neutrino
mass matrix through a type-II seesaw mechanism \cite{Magg:1980ut,
Lazarides:1980nt,Mohapatra:1980yp,Schechter:1980gr}. The fields 
content of the SM is extended with  three right-handed 
neutrinos, one $SU(2)_L$ triplet and a set of SM singlet flavon 
fields. A flavor symmetry $A_4 \times Z_4 \times Z_3$ is considered. 
The type-II seesaw mechanism therefore consists of the conventional 
type-I seesaw contribution ($m_{\nu}^I$) along with the triplet 
contribution ($m_{\nu}^{II}$) to the neutrino mass matrix. 
Here we find the type-I contribution alone can generate the TBM 
mixing pattern, where the charged lepton mass matrix is a diagonal 
one. Then we have shown that the same flavor symmetry allows 
us to have a deviation from the conventional type-I contribution, 
triggered by the $SU(2)_L$ triplet's vev. We have found that this
deviation is sufficient enough to keep $\theta_{13}$ at an
acceptable level\cite{Capozzi:2013csa,Gonzalez-Garcia:2014bfa,
Forero:2014bxa}. We mostly consider the triplet contribution 
to the light neutrino mass is subdominat compared to the 
conventional type-I contribution.

We further assume that apart from the flavons (SM singlets charged under
$A_4$) involved, there is a $A_4$ singlet (as well as SM gauge singlet)
field $S$, which gets a complex vacuum expectation value and thereby 
responsible for spontaneous CP violation\footnote{Earlier it has been 
shown that the idea of spontaneous CP violation \cite{Lee:1973iz} 
can be used to solve strong CP problem \cite{Nelson:1983zb,
Barr:1984qx}. Latter it has been successfully applied on 
models based on $SO(10)$\cite{Harvey:1980je,Harvey:1981hk} 
and other extensions of Standard Model \cite{Branco:1980sz,
Bento:1990wv,Bento:1991ez}.} at high scale 
\cite{Branco:2003rt,Branco:2012vs,Branco:2001pq,Araki:2012hb,Ahn:2013mva,
      Kim:2015etv}. All other flavons have real vevs 
and all the couplings involved are considered to be real.  
It turns out that the magnitude of this complex vacuum expectation value of 
$S$ is responsible for the deviation of TBM by generating nonzero value of
$\theta_{13}$ in the right ballpark. On the other hand, the phase associated 
with it generates the Dirac CP violating phase in the lepton sector. 
So in a way, the triplet contribution provides a unified source for 
CP violation and nonzero $\theta_{13}$. In lepton sector, the other 
possibilities where CP violation can take place, involves complex vev 
of Higgs triplets\cite{Achiman:2004qf,Achiman:2007qz}, or 
when a Higgs bi-doublet (particularly in left-right models) 
gets complex vev\cite{Frank:2004xt} or in a  mixed situation 
\cite{Sahu:2005qm,Chen:2004ww, Chao:2007rm}. However, we will concentrate 
in a situation where a scalar singlet $S$ present in the theory 
gets complex vev as in \cite{Branco:2012vs}. We have also studied the lepton 
asymmetry production through the decay of the heavy triplet involved. 
The decay of the triplet into two leptons contributes to the asymmetry 
where the virtual RH neutrinos are involved in the loop. This process is effective 
when the triplet is lighter than all the RH neutrinos. It turns out that sufficient 
lepton asymmetry can be generated in this way. On the other hand if 
the triplet mass is heavier than the RH neutrino masses, the lightest RH 
neutrino may be responsible for producing lepton asymmetry where the virtual
triplet is contributing in the one loop diagram. 
   
In \cite{Branco:2012vs}, authors investigated a scenario where 
the triplet vevs are the sole contribution to the light neutrino mass and a 
single source of spontaneous CP violation was considered. There, 
it was shown that the low energy CP violating phase and the CP 
violation required for leptogenesis both are governed by the 
argument of the complex vev of that scalar field. The nonzero 
value of $\theta_{13}$ however followed from a perturbative deformation 
of the vev alignment of the flavons involved. Here in our scenario, 
the TBM pattern is realized by the conventional type-I contribution. 
Therefore in the TBM limit, $\theta_{13}$ is zero in our set-up. Also 
there is no CP violating phase in this limit as all the flavons 
involved in $m_{\nu}^I$ are carrying real vevs, and hence no 
lepton asymmetry as well. Now once the triplet contribution 
($m_{\nu}^{II}$) is switched on, not only the $\theta_{13}$ , 
but also the leptonic CP violation turn out to be nonzero. For generating 
lepton asymmetry, two triplets were essential in \cite{Branco:2012vs}, 
while we could explain the lepton asymmetry by a single triplet along with the 
presence of RH neutrinos. In this case, the RH neutrinos are heavier compared to
the mass of the triplet involved.

The paper is organized as follows. In section \ref{sec2}, we provide the status of the 
neutrino mixing and the mass squared differences. Then in section \ref{sec3}, we 
describe the set-up of the model followed by constraining the parameter 
space of the framework from neutrino masses and mixing in section \ref{nucons}. 
In section \ref{sec:lep}, we describe how one can obtain lepton asymmetry out of this 
construction. Finally we conclude in section \ref{conlc}.


\section{Status of Neutrino Masses and Mixing:}\label{sec2}

Here we summarize the neutrino mixing parameters and their present status.
The neutrino mass matrix $m_{\nu}$, in general, can be diagonalized by the 
$U_{PMNS}$ matrix (in the basis where charged lepton mass matrix is diagonal) as 
$m_{\nu} = U^*_{PMNS} {\rm diag}(m_1, m_2, m_3)U^{\dagger}_{PMNS}$, 
where $m_1, m_2, m_3$ are the real mass eigenvalues for light neutrinos. 
The standard parametrization  of the $U_{PMNS}$ matrix \cite{Agashe:2014kda} is 
given by
{\small
\begin{eqnarray}
U_{PMNS}=\left(
\begin{array}{ccc} 
    c_{12}c_{13}                                &   s_{12}c_{13}
               & s_{13}e^{-i\delta}\\
    -s_{12}c_{23}-c_{12}s_{13}s_{23}e^{i\delta} &
c_{12}c_{23}-s_{12}s_{13}s_{23}e^{i\delta}    &   c_{13}s_{23} \\
     s_{12}s_{23}-c_{12}s_{13}c_{23}e^{i\delta} &
-c_{12}s_{23}-s_{12}s_{13}c_{23}e^{i\delta}    &   c_{13}c_{23}
\end{array}
\right)
\left(
\begin{array}{ccc}
  1 &0 &0\\
  0 &e^{i\alpha_{21}/2} &0\\
  0 &0 &e^{i\alpha_{31}/2}
\end{array}
\right),
\label{upmns}
\end{eqnarray}
}
where $c_{ij}=\cos\theta_{ij}$, $s_{ij}=\sin\theta_{ij}$, 
the angles $\theta_{ij} = [0, \pi/2]$, $\delta=[0, 2\pi]$ is the
CP-violating Dirac phase while $\alpha_{21}$ and $\alpha_{31}$ are the two
CP-violating Majorana phases. 
\begin{table}[h]\centering
\resizebox{12.9cm}{!}{
  \begin{tabular}{|c|c|c|c|c|}
    \hline
    Oscillation parameters&best fit & $1\sigma$ range&  3$\sigma$ range
    \\
    \hline\hline
    $\Delta m^2_{21}$ [$10^{-5}\hspace{.1cm} \eVq$]
   &7.60 & 7.42--7.79   & 7.11--8.18 \\
   \hline
    $|\Delta m^2_{31}|$ [$10^{-3}\hspace{.1cm}\eVq$] 
    &
    \begin{tabular}{c}
      2.48  (NH)\\
      2.38 (IH)\\
    \end{tabular}
    &
    \begin{tabular}{c}
      $2.41-2.53$  \\
      $2.32-2.43$ \\
    \end{tabular}
    &
    \begin{tabular}{c}
      $2.30-2.65$\\
      $2.20-2.54$
    \end{tabular}
    \\
    \hline
    $\sin^2\theta_{12}$
    & 0.323& 0.307--0.339 & 0.278--0.375\\
    \hline
    $\sin^2\theta_{23}$
      &
    \begin{tabular}{c}
      0.567 (NH)\\
      0.573 (IH)\\
    \end{tabular}
    &
    \begin{tabular}{c}
      0.439--0.599\\
      0.530--0.598\\
    \end{tabular}
    &
    \begin{tabular}{c}
      0.392--0.643\\ 
      0.403--0.640 
    \end{tabular} \\
    \hline
    $\sin^2\theta_{13}$
     &
    \begin{tabular}{c}
      0.0234 (NH) \\
      0.0240 (IH)
    \end{tabular}
    &
    \begin{tabular}{c}
      0.0214--0.0254  \\
      0.0221--0.0259 
    \end{tabular}
    &
   \begin{tabular}{c}
      0.0177--0.0294 \\
      0.0183--0.0297
    \end{tabular} \\
    \hline
       \hline
     \end{tabular}
}
     \caption{ \label{tab1} {\small Summary of neutrino oscillation parameters
     for normal and inverted neutrino mass hierarchies from the analysis of
     \cite{Forero:2014bxa}.}}
\end{table}
The mixing angles $\theta_{12}$,
$\theta_{23}$ and the two mass-squared differences $\Delta m^2_{12} (\equiv m^2_2 - m^2_1), 
~ \Delta m^2_{31} (\equiv m^2_3 - m^2_1)$ have been well measured at
several neutrino oscillation experiments \cite{NuExpt}. Recently the other mixing angle 
$\theta_{13}$ is also reported to be of sizable magnitude \cite{Abe:2011fz, DayaBay, 
Ahn:2012nd, Abe:2013hdq}.  Very recently, we start to get hint for nonzero Dirac CP
phase\cite{Capozzi:2013csa,Forero:2014bxa,Gonzalez-Garcia:2014bfa,Abe:2013hdq}. 
From the updated global analysis \cite{Forero:2014bxa}
involving all the data from neutrino experiments, the 1$\sigma$ and
3$\sigma$ ranges of mixing angles and the mass-squared differences are mentioned
(NH and IH stand for the normal and inverted mass hierarchies respectively) in
Table \ref{tab1}.
The result by Planck\cite{Ade:2013zuv} from the analysis of cosmic microwave background
(CMB) also sets an upper limit on the sum of the three neutrino masses
as given by, $\Sigma_{i} m_{\nu_i} < 0.23$ eV. The result from 
neutrinoless double beta decay by KamLAND-Zen\cite{Asakura:2014lma} and 
EXO-200 \cite{Albert:2014awa} indicates a limit on the effective neutrino
mass parameter $|m_{ee}|$ as, $|m_{ee}| < (0.14-0.28)$ eV at 90$\%$ CL
and $|m_{ee}| < (0.19 - 0.45)$ eV at 90$\%$ CL respectively.


\section{The Model }\label{sec3}

Our starting point is the conventional type-I seesaw mechanism to explain 
the smallness of light neutrino masses which further predicts a 
tribimaximal mixing (TBM) pattern in the lepton sector. For this part, 
we use the original AF model \cite{Altarelli:2005yx}
by introducing a discrete $A_4$ symmetry and $A_4$ triplet flavon fields
$\phi_{\sm}, \phi_{\T}$ along with a singlet $\xi$ field. Of course three
right handed neutrinos ($N_{\R}$) are also incorporated. 
In addition, we include a $SU(2)_L$ triplet  field ($\Delta$) with
hypercharge unity, the vev
of which produces an additional contribution (hereafter called the triplet
contribution) to the light neutrino mass. So our set-up basically involves
a general type-II seesaw, 
\begin{equation}
m_{\nu} = m_{\nu}^{II} + m_{\nu}^{I} = m_{\nu}^{II} - m_{\D}^T M_{\R}^{-1} m_{\D},
\end{equation}
where $m_{\nu}^{I} $ is the typical type-I term and $m_{\nu}^{II}$ is the
triplet contribution. To realize both, the relevant Lagrangian for generation
of $m_{\nu}$ can be written as,
\begin{equation}
-\mathcal{L}=Y_{\D}\bar{L}\widetilde{H}N_{\R}
              + \frac{1}{2}M_{\R}\overline{N_{\R}^c}N_{\R}
              +(Y_{\mb})_{ij}L^T_{i}C\Delta L_{j},
\end{equation}
so that $m_{\nu}^{II} =2 Y_{\mb} u_{\mb}$ and $m_{\D} = Y_{\D} v$, where $u_{\mb}$ 
and $v$ are the vevs of the triplet $\Delta$ and SM Higgs doublet ($H$)  
respectively. $Y_{\D}$ and $Y_{\mb}$ correspond to the Yukawa matrices for
the Dirac mass and triplet terms respectively, the flavor structure of 
which are solely determined by the discrete symmetries imposed on the fields
involved in the model. $M_{\R}$ is the Majorana mass of the RH neutrinos. In the
following subsection, we discuss in detail how the flavor structure of
$Y_{\D}, Y_{\mb}$ and $M_{\R}$ are generated with the flavon fields. 
A discrete symmetry $Z_4 \times Z_3$ is also present in our model and two
other SM singlet fields $\xi'$ and $ S$ are introduced. These additional  
fields and the discrete symmetries considered play crucial role in realizing a typical 
structure of the triplet contribution to the light neutrino mass matrix as 
we will see below. Among all these scalar fields present, only the $S$ field
is assumed to have a complex vev while all other vevs are real. The framework
is based on the SM gauge group extended with the $A_4 \times Z_4 \times Z_3$
symmetry. The field contents and charges under the symmetries imposed are 
provided in Table \ref{teb2}. 
\begin{table}[h]
\centering
\resizebox{15cm}{!}{%
\begin{tabular}{|c|cccccccccccc|}
\hline
 Field & $e_{\R}$ & $\mu_{\R}$ & $\tau_{\R}$  & $L$ & $N_{\R}$ & $H$ & 
 $\Delta$ & $\phi_{\sm}$ & $\phi_{\T}$ &  $\xi$ & $\xi'$ & $S$ \\
\hline
$A_{4}$ & 1 & $1''$ & $1'$ &  3 & 3 & 1 & 1 & 3 & 3 & 1&$1'$ &1 \\
\hline
$Z_{4}$ &$-1$ & $-1$ & $-1$ & $i$ & $i$ & $1$ &
$-i$ & $-1$ & $-i$ & $-1$ & $i$ &$-1$\\
\hline
$Z_{3}$ &$\omega$ & $\omega$ & $\omega$ & $\omega$ & $\omega$ & $1$ &
$\omega^2$ & $\omega$ & $1$ & $\omega$ & $\omega^2$ &$1$ \\
\hline
\end{tabular}
}\
\caption{\label{teb2} {\small Fields content and transformation properties
under the symmetries imposed on the model.}}
\end{table}

With the above fields content, the charged lepton Lagrangian is described
by, 
\begin{equation}
 \mathcal{L}_l =  \frac{y_e}{\Lambda}(\bar{L}\phi_{\T})H e_{\R}
+\frac{y_{\mu}}{\Lambda}(\bar{L}\phi_{\T})'H\mu_{\R}+ 
\frac{y_{\tau}}{\Lambda}(\bar{L}\phi_{\T})''H\tau_{\R},
\end{equation}
to the leading order, where $\Lambda$ is the cut-off scale of the theory 
and $y_e, y_{\mu}$ and $ y_{\tau}$ are the respective coupling constants. Terms in 
the first parenthesis represent products of two $A_4$ triplets, which 
further contracts with $A_4$ singlets $1$, $1''$ and $1'$ corresponding to 
$e_{\R}, \mu_{\R}$ and $\tau_{\R}$ respectively to make a true singlet 
under $A_4$.  Once the flavons $\phi_{\sm}$ and $\phi_{\T}$ get the vevs 
along a suitable direction as $(u_{\sm}, u_{\sm}, u_{\sm})$ and $ (u_{\T},0, 0)$ 
respectively\footnote{The typical vev alignments of $\phi_{\sm}$ and 
$\phi_{\T}$ are assumed here. We expect the minimization of the
potential involving $\phi_{\sm}$ and $\phi_{\T}$ can produce this by
proper tuning of the parameters involved in the potential. However the
very details of it are beyond the scope of this paper.}, it 
leads to a diagonal mass matrix for charged leptons, once the Higgs vev 
$v$ is inserted. Below we will first summarize how the TBM mixing is 
achieved followed by the triplet contribution in the next subsection. 
The requirement of introducing SM singlet fields will be explained 
subsequently while discussing the flavor structure of neutrino mass 
matrix in detail. 
  
\subsection {Type-I Seesaw and Tribimaximal Mixing}

The relevant Lagrangian for the type-I seesaw in the neutrino sector is
given by, 
\begin{equation}\label{l1}
 \mathcal{L_I} =  y\bar{L}\widetilde{H}N_{\R}
+x_{\A}\xi{\overline{N_{\R}^c}N_{\R}}+x_{\B}\phi_{\sm}
\overline{N_{\R}^c}N_{\R},
\end{equation}
where $y, x_A$ and $x_B$ are the coupling constants. 
After the $\xi $ and $\phi_{\sm}$ fields get vevs and the electroweak vev
$v$ is included, it yields the following flavor structure for Dirac ($m_{\D}$)
and Majorana ($M_R$) mass matrices, 
\begin{eqnarray}
m_{\D}=Y_{\D}v=y v  \left(
\begin{array}{ccc}
      1 &0 &0\\
       0 &0 &1\\
       0 &1 &0
\end{array}
\right)
\hspace{.3cm}{\rm and}\hspace{.3cm}
M_{\R} = \left(
\begin{array}{ccc}
      a+2b/3   &-b/3    &-b/3\\
       -b/3    &2b/3    &a-b/3\\
       -b/3    &a-b/3   &2b/3
\end{array}
\right),
\label{MDMR}
\end{eqnarray}
with $a = 2x_{\A}\langle \xi \rangle=2x_{\A}u_{\xx}, b = 2x_{\B} u_{\sm}$.
The $A_4$ multiplication rules that results to this flavor structure can 
be found in \cite{Karmakar:2014dva}.  Therefore the contribution toward
light neutrino mass that results from the type-I seesaw mechanism is 
found to be, 
\begin{eqnarray}\label{T1}
 m_{\nu}^I &=& - m_{\D}^TM_{\R}^{-1}m_{\D}\nonumber\\
                   &=& -y^2v^2\left(
\begin{array}{ccc}
 \frac{3 a+b}{3 a(a+b)} & \frac{b}{3 a(a+b)} & \frac{b}{3 a(a+b)} \\
 \frac{b}{3 a(a+b)} & -\frac{b (2 a+b)}{3a  \left(a^2-b^2\right)} & 
 \frac{3 a^2+a b-b^2}{3a  \left(a^2-b^2\right)} \\
 \frac{b}{3 a(a+b)} & \frac{3 a^2+a b-b^2}{3a  \left(a^2-b^2\right)}
 & -\frac{b (2 a+b)}{3a  \left(a^2-b^2\right)}
\end{array}
\right).
\end{eqnarray}
Note that this form of $m_{\nu}^I$ indicates that the corresponding 
diagonalizing matrix would be nothing but the TBM mixing matrix of 
the form \cite{Harrison:1999cf}
 \begin{eqnarray}\label{utb}
 U_{TB}=\left(
\begin{array}{ccc}
 \sqrt{\frac{2}{3}} & \frac{1}{\sqrt{3}} & 0 \\
 -\frac{1}{\sqrt{6}} & \frac{1}{\sqrt{3}} & -\frac{1}{\sqrt{2}} \\
 -\frac{1}{\sqrt{6}} & \frac{1}{\sqrt{3}} & \frac{1}{\sqrt{2}}
\end{array}
\right).
\end{eqnarray}

As a characteristic of typical $A_4$  generated structure, the RH 
neutrinos mass matrix is as well diagonalized by the $U_{TB}$. In order 
to achieve the real and positive mass eigenvalues, the corresponding 
rotation $U_{\R}$ is provided on $M_{\R}$ as $U_{\R}^T M_{\R}U_{\R}=
M_{\R}^{\rm diag}={\rm diag}(a+b,a,a-b)$ with $U_{\R}=U_{TB}{\rm diag}
(1,1,e^{-i\pi/2})$ once $a>b$ is considered. On the other hand for $a<
b$; through $U_{\R}=U_{TB}$ itself, the real and positive eigenvalues 
of $M_{\R}$ [$M_{\R}^{\rm diag}={\rm diag}(a+b,a,b-a)$] can be obtained.
This would be useful when we will consider the decay of the RH neutrinos for 
leptogenesis in Section \ref{sec:lep}.

\subsection{Triplet Contribution and Type-II seesaw}

The leading order Lagrangian invariant under the symmetries 
imposed, that describes the triplet contribution to the light neutrino
mass matrix ($m_{\nu}^{II}$), is given by, 
\begin{equation} \label{ld}
 \mathcal{L_{II}} = \frac{1}{\Lambda^2}\Delta{L^T}L(x_1S+x_1'S^{*})\xi',
\end{equation}
where $x_1$ and $x_1'$ are the couplings involved.  Here $\xi'$ develops
a vev $u_{\x}$ and the singlet $S$ is having a complex vev $\langle S \rangle
= v_{\sm} e^{i\alpha_{\sm}}$. As we have mentioned before, the vev of $S$
provides the unique source of CP violation as all other vevs and couplings
are assumed to be real. CP is therefore assumed
to be conserved in all the terms involved in the Lagrangian. Similar to
\cite{Branco:2012vs}, CP is spontaneously broken by the complex vev
of the $S$ field. After plugging all these vevs, the above Lagrangian in 
Eq.(\ref{ld}) contributes to the following Yukawa matrix for the triplet
$\Delta$ as given by, 
\begin{eqnarray}
 Y_{\Delta} =h  \left(
\begin{array}{ccc}
0 & 0& 1 \\
0 & 1 & 0 \\
1 &0 &0
\end{array}
\right), \hspace{.3cm} 
h =\frac{1}{\Lambda^2}u_{\x}v_{\sm}(x_1e^{i\alpha_{\sm}}+x_1'e^{-i\alpha_{\sm}}).
\label{YDelta}
\end{eqnarray}
This specific structure follows from the $A_4$ charge assignments of various 
fields present in Eq.(\ref{ld}) and is instrumental in providing nonzero 
$\theta_{13}$ as we will see shortly.

Before discussing the vev of the $\Delta$ field, let us describe the complete
scalar potential $V$, including the triplet $\Delta$ obeying
the symmetries imposed, is given by, 
\begin{equation}
 V=V_S+V_H+V_{\Delta}+V_{SH}+V_{S\Delta}+V_{\Delta H},
\end{equation}
where
\begin{eqnarray}\label{sp}
 V_{S}&=&\mu_{\sm}^2(S^2+S^{*2})+m_{\sm}^2S^*S+\lambda_1(S^4+S^{*4})
        +\lambda_2 S^*S(S^2+S^{*2})+\lambda_3 (S^*S)^2,\nonumber\\
 V_H&=&m^2_HH^{\dagger}H+\lambda_4 (H^{\dagger}H)^2,\nonumber\\\nonumber
 V_{\Delta}&=&{M_{\mb}^2} {\rm Tr}(\Delta^{\dagger}\Delta)+
 \lambda_5 [{\rm Tr}(\Delta^{\dagger}\Delta)]^2,\\\nonumber
 V_{SH}&=&\lambda_6 (S^*S)H^{\dagger}H+\lambda_7
(S^2+S^{*2})(H^{\dagger}H),\nonumber\\
 V_{S\Delta}&=&{\rm Tr}(\Delta^{\dagger}\Delta)
 [\lambda_8 (S^2+S^{*2})+\lambda_9 S^*S],\nonumber\\
 V_{\Delta H}&=&\lambda_{10}(H^{\dagger}H){\rm Tr}(\Delta^{\dagger}\Delta)
 +\lambda_{11}(H^{\dagger}\Delta^{\dagger}\Delta H) + 
 \left(-\frac{\mu}{\Lambda}\widetilde{H}^T\Delta\widetilde{H}\phi_{\sm}\phi_{\T}
            +h.c.\right).
\end{eqnarray}
The above potential contains several dimensionful (denoted by $\mu_{\sm}
, m_{\sm,\hh}, M_{\mb}$) and dimensionless parameters (as $\lambda_{i=1,2,..11}
\hspace{.1cm} {\rm and} \hspace{.1cm} \mu$), which are all considered to
be real. Similar to \cite{Branco:2012vs}, here also it can be shown that 
the $S$ field gets a complex vev for a choice of parameters involved in 
$V_S$ as $m^2_S < 0,\mu_{\sm} \simeq 0$ and $\lambda_3>2\lambda_1>0$.
However contrary to
\cite{Branco:2012vs}, here we have only a single triplet field $\Delta$.
Once the $\phi_{\sm}, \phi_{\T}$ get vevs, the last term of $V_{\Delta H}$ 
results into an effective $\Delta H H$ interaction which would be important for
leptogenesis. The vev of the triplet $\Delta$ is obtained by minimizing
the relevant terms\footnote{We consider couplings 
$\lambda_{8,9} \ll 1$.} from $V$
after plugging the vevs of the flavons and is given by 
\begin{equation}
 \langle\Delta^0\rangle \equiv u_{\Delta} =  \eta \frac{v^2}{M_{\mb}^2}
 ~~{\rm and}~~
 \eta = \frac{\mu}{\Lambda}  u_{\sm} u_{\T}. 
\label{D1D2}
\end{equation}

Using Eqs.(\ref{YDelta}) and (\ref{D1D2}), the triplet contribution 
to the light neutrino mass matrix follows from the Lagrangian 
$\mathcal{L_{II}}$ as   
\begin{eqnarray}
m_{\nu}^{II} =  \left(
\begin{array}{ccc}
      0   & 0    & d\\
      0   & d    & 0\\
      d   & 0    &0
\end{array}
\right),\label{d}
\end{eqnarray}  where
\begin{eqnarray}\label{modd}
 d= 2h u_{\Delta}
   = 2  h \eta \frac{v^2}{M^2_{\Delta}} .
\end{eqnarray}
Note that only the triplet contribution ($d$) involves the phase due to
the involvement of $\langle S \rangle$ in $h$, while the entire type-I 
contribution $m_{\nu}^I$ remains real. Therefore the term $d$ serves as
the unique source of generating all the CP-violating phases involved in
neutrino as well as in lepton mixing. This will be clear once we discuss
the neutrino mixing in the  subsequent section.  
Now we can write down the entire contribution to the light neutrino mass as,  
\begin{eqnarray}\label{m1p2}
 m_{\nu}&=& -y^2v^2\left(
\begin{array}{ccc}
 \frac{3 a+b}{3 a(a+b)} & \frac{b}{3 a(a+b)} & \frac{b}{3 a(a+b)} \\
 \frac{b}{3 a(a+b)} & -\frac{b (2 a+b)}{3a  \left(a^2-b^2\right)} & \frac{3 
a^2+a b-b^2}{3a  \left(a^2-b^2\right)} \\
 \frac{b}{3 a(a+b)} & \frac{3 a^2+a b-b^2}{3a  \left(a^2-b^2\right)} & -\frac{b (2 
a+b)}{3a  \left(a^2-b^2\right)}
\end{array}
\right)+
\left(
\begin{array}{ccc}
0 & 0& d \\
0 & d & 0 \\
d &0 &0
\end{array}
\right).
\end{eqnarray}

\section{Constraining parameters from neutrino mixing}\label{nucons}
In this section, we discuss how the neutrino masses and mixing can be obtained
from the $m_{\nu}$ mentioned above. Keeping in mind
that $m_{\nu}^I$ can be diagonalized by $U_{TB}$, we first perform a 
rotation by $U_{TB}$ on the explicit form of the light neutrino mass matrix 
obtained in Eq.(\ref{m1p2}) and the rotated $m_{\nu}$ is found to be
\begin{eqnarray}
m'_{\nu}=U_{TB}^Tm_{\nu}U_{TB}
&=&\left(
\begin{array}{ccc}
 -\frac{a d+b d+2 v^2 y^2}{2 (a+b)} & 0 & \frac{\sqrt{3} d}{2} \\
 0 & d-\frac{v^2 y^2}{a} & 0 \\
 \frac{\sqrt{3} d}{2} & 0 & \frac{a d-b d+2 v^2 y^2}{2 (a-b)}
\end{array}
\right),\\
&=&\left(\begin{array}{ccc}\label{mnupl1}
 -\frac{d}{2}-\frac{k}{ (1+\alpha)} &   0   & \frac{\sqrt{3} d}{2} \\
          0                         &  d-k  &       0              \\
 \frac{\sqrt{3} d}{2}               &   0   & \frac{d}{2}+\frac{k}{ (1-\alpha)}
\end{array}
\right).
\end{eqnarray}
Here, we define the parameters $\alpha=b/a$ and $k={v^2 y^2}/{a}$ which are 
real and positive as part of the type-I contribution. We note that a further
rotation by $U_1$ (another unitary matrix) in the 13 plane in required to 
diagonalize the light neutrino mass matrix, $U^T_{1}m'_{\nu}U_1=m^{\rm diag}_{\nu}$.
With a form of $U_1$ as  
\begin{eqnarray}\label{u1}
U_{1}=\left(
\begin{array}{ccc}
 \cos\theta               & 0 & \sin\theta{e^{-i\psi}} \\
     0                    & 1 &            0 \\
 -\sin\theta{e^{i\psi}} & 0 &        \cos\theta
\end{array}
\right),
\end{eqnarray}
we have, $(U_{TB}U_1)^Tm_{\nu}U_{TB}U_1 = {\rm diag} (m_1e^{i\gamma_1},
m_2e^{i\gamma_2},m_3e^{i\gamma_3})$, where $m_{i=1,2,3}$ are the real 
and positive eigenvalues and $\gamma_{i=1,2,3}$ are the phases 
associated to these mass eigenvalues. We can therefore extract
the neutrino mixing matrix $U_{\nu}$ as, 
\begin{eqnarray}\label{unu}
 U_{\nu} =U_{TB}U_1U_m= \left(
\begin{array}{ccc}
 \sqrt{\frac{2}{3}}\cos\theta & \frac{1}{\sqrt{3}} & \sqrt{\frac{2}{3}} e^{-i \psi } \sin\theta \\
 -\frac{\cos\theta}{\sqrt{6}}+\frac{e^{i \psi } \sin\theta}{\sqrt{2}} & \frac{1}{\sqrt{3}}
 & -\frac{\cos\theta}{\sqrt{2}}-\frac{e^{-i \psi }\sin\theta}{\sqrt{6}} \\
 -\frac{\cos\theta}{\sqrt{6}}-\frac{e^{i \psi } \sin\theta}{\sqrt{2}} & 
 \frac{1}{\sqrt{3}} & \frac{\cos\theta}{\sqrt{2}}-\frac{e^{-i \psi }\sin\theta}{\sqrt{6}}
\end{array}
\right)U_{m}
\end{eqnarray}
where $U_{m}={\rm diag} (1,e^{i\alpha_{21}/2},e^{i\alpha_{31}/2})$ is 
the Majorana phase matrix with $\alpha_{21}=(\gamma_1-\gamma_2)$ and 
$\alpha_{31}=(\gamma_1-\gamma_3)$, one common phase being irrelevant.
As the charged lepton mass matrix is a diagonal one, we can  now 
compare this $U_{\nu}$ with the standard  parametrization of lepton
mixing matrix $U_{PMNS}$. The $U_{PMNS}$ 
is therefore given by $U_{PMNS}=U_{\rm P}U_{\nu}$, where we need to multiply the 
$U_{\nu}$ matrix by a diagonal phase matrix $U_{\rm P}$ \cite{King:2011zj}
from left so that the $U_{PMNS}$ excluding the Majorana phase matrix,  can take the 
standard form where 22 and 33 elements are real as in Eq.(\ref{upmns}). Hence we
obtain the usual (in $A_4$ models) correlation \cite{Altarelli:2012ss}
between the angles and CP violating Dirac phase $\delta$ as given by 
\begin{eqnarray}\label{ang}
\sin\theta_{13}=\sqrt{\frac{2}{3}}\left|\sin\theta\right|, \hspace{.2cm}
\sin^2\theta_{12}=\frac{1}{3(1-\sin^2\theta_{13})},\\
 \sin^2\theta_{23}=\frac{1}{2}
+\frac{1}{\sqrt{2}}\sin\theta_{13}\cos\delta, \hspace{.2cm} \delta={\rm arg}[(U_1)_{13}]
\label{ang22}. 
\end{eqnarray}

The angle $\theta$ and phase $\psi$ associated with $U_1$ can now be linked
with the parameters involved in $m_{\nu}$. For this we first rewrite the
triplet contribution $d$ as $d = \vert d \vert e^{i\phi_d}$ and define a 
parameter $\beta = {\vert d \vert}/k$ (hence $\beta$ is real). 
This parameter indicates the relative size of the triplet contribution to the
type-I contribution when $\alpha \leq 1$. As $U_1$ diagonalizes the $m'_{\nu}$
matrix, after some involved algebra, we finally get, 
\begin{equation}
\tan{2\theta} = \frac{\sqrt{3}}{\alpha} 
                 \frac{
                 \left [ 1 - \left( 1- \alpha^2 \right) \cos^2{\phi_d} \right]^{1/2}
                 }
                 {\frac{2}{\beta \left( 1-\alpha^2\right)} + \cos{\phi_d}}
                 ~~{\rm and} ~~ 
\tan{\psi} = (\tan{\phi_d})/\alpha.
\label{theta-psi}
\end{equation}
$\sin\theta$ may take positive or negative value depending on the choices of
$\alpha, \beta$ as evident from the first relation in Eq.(\ref{theta-psi}).
For $\sin\theta>0$, we find $\delta=\psi$ using $\delta={\rm arg}[(U_1)_{13}$ and 
the second relation of Eq.(\ref{theta-psi}).On the other hand for $\sin\theta<0$;
$\delta$ and $\psi$ are related by $\delta=\psi \pm \pi$. Therefore in both
these cases we obtain $\tan{\psi} = \tan{\delta}$ and hence 
\begin{equation}
\tan\delta  =  (\tan{\phi_d})/\alpha.
\label{delta-phid}
\end{equation}
In our set-up, the source 
of this CP-violating Dirac phase $\delta$ is through the phase $\alpha_{\sm}$
associated with $\langle S \rangle$. Note that $\tan\delta$ is related with 
$\tan\phi_d$ and $\alpha$ as seen from Eq.(\ref{delta-phid}). Now from the 
relation $d=|d|e^{i\phi_d}$ and using Eq.(\ref{YDelta}) and (\ref{modd}), we
obtain $\phi_d$ satisfying  
\begin{equation}\label{phidal}
 \tan\phi_d=\frac{ (x_1-x'_1)}
                 {(x_1+x'_1)}\tan\alpha_{\sm},
\end{equation}
where $x_1$ and $x_1'$ are the coupling involved in Eq.(\ref{ld}).

As seen from Eqs.(\ref{ang}) and (\ref{theta-psi}),
we conclude that the $U_{PMNS}$ parameters $\theta_{13}$ and $\delta$ depend on 
the model parameters $\alpha$, $\beta$ and $\phi_d$. Note that we expect terms 
$a$ and $b$ ($\alpha=b/a$) to be of similar order of magnitude as both originated
from the tree level Lagrangian (see Eqs.(\ref{l1}) and (\ref{MDMR})). We categorize
$\alpha<1$  as case A, while $\alpha>1$ is with case B. The other parameter 
$\beta$ basically represents the relative order of magnitude 
between the triplet contribution ($\vert d \vert$) and the type-I contribution 
($v^2 y^2/ a$). Our framework produces the TBM mixing pattern to be generated 
solely by type-I seesaw and triplet contribution is present mainly to correct 
for the angle $\theta_{13}$ which is small compared to the other mixing angles. 
Therefore we consider that the triplet contribution is preferably the sub-dominant or 
at most comparable one. Therefore we expect the parameter $\beta$ to be less than one. 
\begin{figure}[!h]
\begin{center}
\includegraphics[scale=0.7]{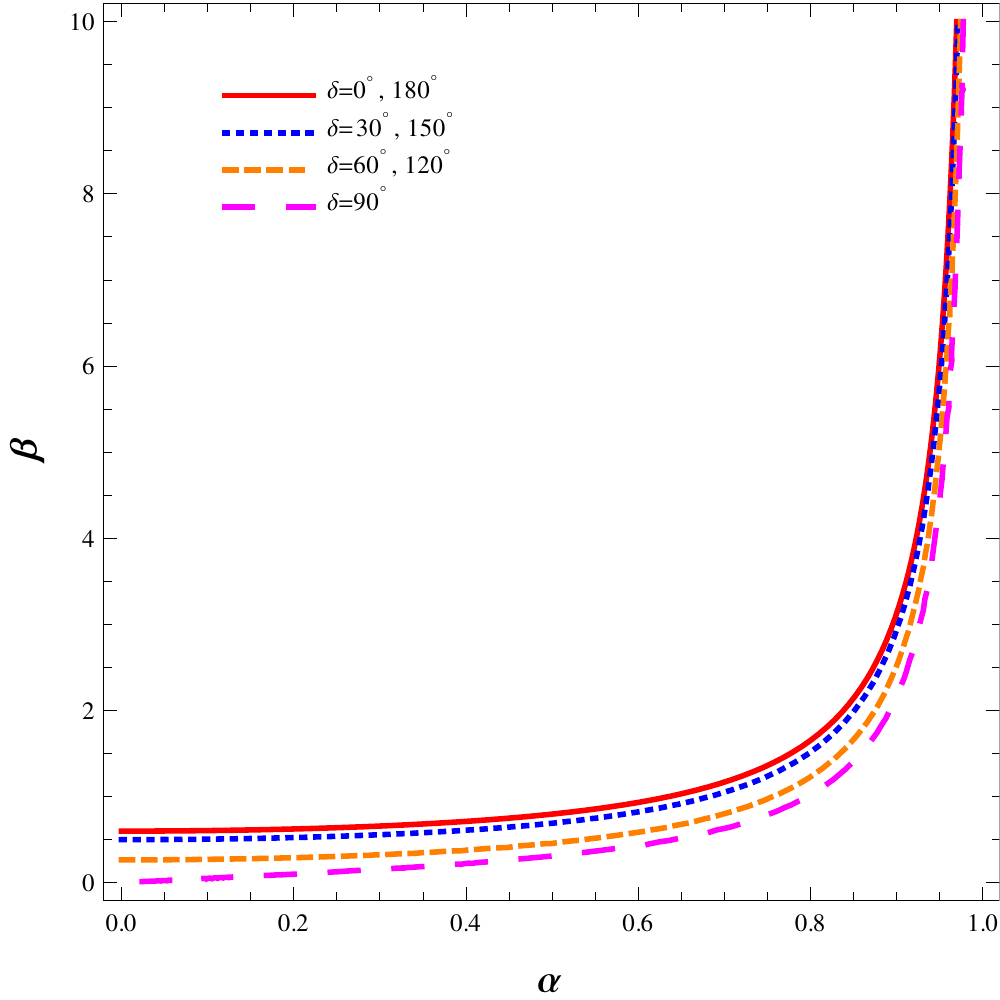}
\includegraphics[scale=0.7]{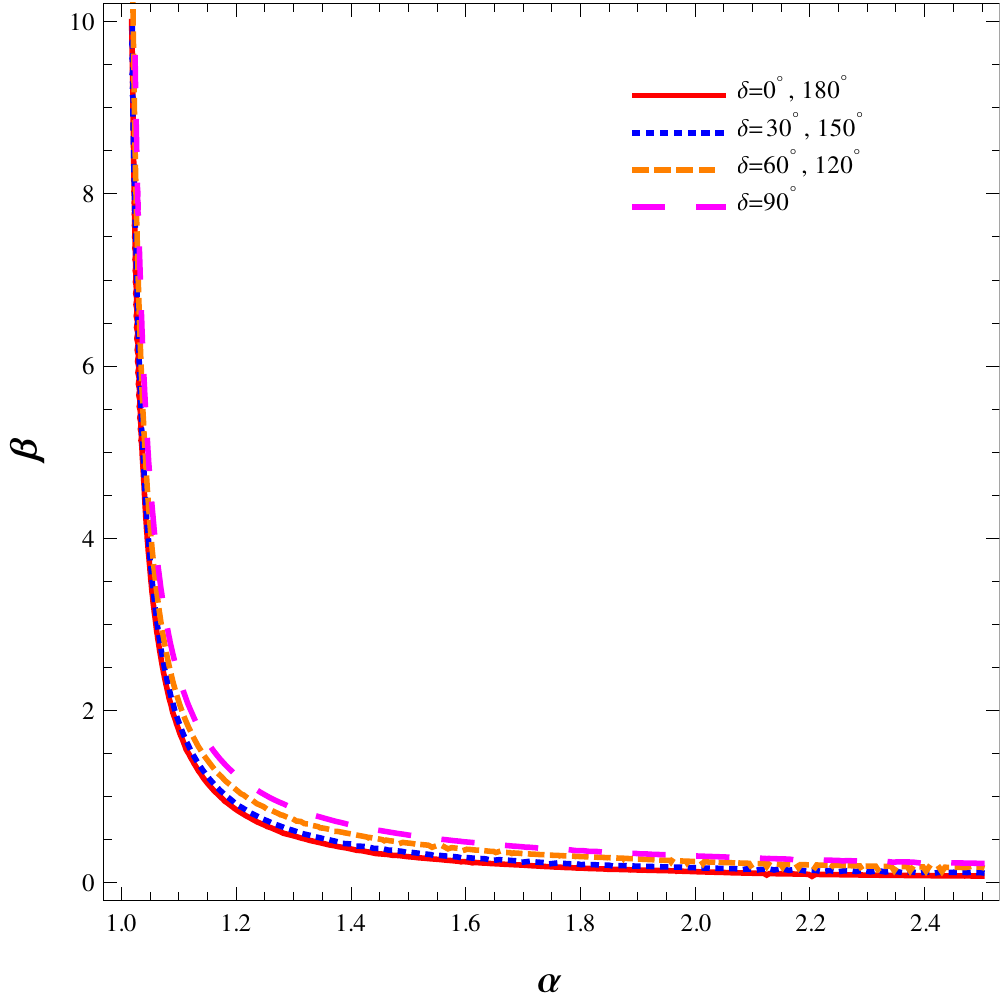}
\caption{{\small Contour plots for $\sin^2\theta_{13}=0.0234$ in the $\alpha-\beta$
         plane for various choices of $\delta$ as indicated inside the figure. 
         Left panel is for (A) $\alpha<1$ and right panel is with (B)
         $\alpha>1$. 
         }}
\label{s13l}
\end{center}
\end{figure}
Although we discuss what happens when $\beta > 1$ in some cases, we will restrict 
ourselves with $\beta < 1$ for the most of the analyses involved later in this work. 
In Fig.\ref{s13l} left panel, we study the variation of $\alpha$ and
$\beta$ in order to achieve the best fit value of $\sin^2 \theta_{13} =
0.0234$ \cite{Forero:2014bxa} while different values of $\delta$ are 
considered. In producing these plots, we have replaced the $\phi_d$ dependence
in terms of $\alpha$ and $\delta$ by employing the second equation
in Eq.(\ref{theta-psi}) as $\psi = \delta$. Similarly in the right panel of Fig.\ref{s13l},
contour plots for $\sin^2 \theta_{13} = 0.0234$ are depicted for $\alpha>1$
with different values of $\delta$. We find a typical contour plot for $\sin^2{\theta_{13}}$ with a 
specific $\delta$ value coinsides with the one with other $\delta$ values obtained from 
$|\pi - \delta|$. For example, one particular contour plot for $\delta = 30^\circ$ is repeated for 
$\delta = 150^\circ, 210^\circ, 330^\circ$.
 
Diagonalizing $m_{\nu}'$ in Eq.({\ref{mnupl1}}), the light neutrino masses
turn out to be, 
\begin{eqnarray}
 m_1&=&k\left[\left( \frac{\alpha}{\pm(1-\alpha^2)}  - \frac{p}{k} \right)^2
                   +\left(\frac{q}{k}\right)^2\right]^{1/2},\label{m1}\\
m_2&=&k\left[1+\beta^2-2\beta\cos\phi_d\right]^{1/2},\label{m2}  \\
m_3&=&k\left[\left(\frac{\alpha}{\pm(1-\alpha^2)} + \frac{p}{k} \right)^2
                  +\left(\frac{q}{k}\right)^2\right]^{1/2}\label{m3},
\end{eqnarray}
where $p$ and $q$ are defined as,
\begin{eqnarray}
\left( \frac{p}{k}\right)^2=\frac{1}{2}\left( \frac{A}{k^2}
        +\sqrt{\frac{A^2}{k^4}+\frac{B^2}{k^4}}\right) & \rm{,} &
 \left(\frac{q}{k}\right)^2=\frac{1}{2}\left(-\frac{A}{k^2}
        +\sqrt{\frac{A^2}{k^4}+\frac{B^2}{k^4}}\right); \\
\frac{A}{k^2}=  \beta^2\cos 2\phi_d
                 +\beta\frac{\cos\phi_d}{1-\alpha^2} + 
                 \frac{1}{(1-\alpha^2)^2} & \rm{,}&
 \frac{B}{k^2}=\beta^2\sin 2\phi_d
                 +\beta\frac{\sin\phi_d}{1-\alpha^2}.                 
\end{eqnarray}
The `$+$' sign in the expression of $m_1$ and $m_3$ is for $\alpha<1$
(case A) where the `$-$' sign is associated with $\alpha>1$ (case B).
The Majorana phases in $U_m$ (see Eq.(\ref{unu})) are found to be 
\begin{eqnarray}
\alpha_{21}&=& \tan^{-1}\left[
                    \frac{q/k}{p/k \pm \frac{\alpha}{(\alpha^2 - 1)}}
                    \right]\label{majo1}-\tan^{-1}\left[
              \frac{\beta\sin\phi_d}{\beta\cos\phi_d-1}
                     \right]\label{majo2} ,\\
\alpha_{31}&=&\pi+\tan^{-1}\left[
                    \frac{q/k}{p/k \pm \frac{\alpha}{(\alpha^2 - 1)}}
                    \right]\label{majo1}-\tan^{-1}\left[
                  \frac{q/k}{p/k \pm \frac{\alpha}{(1-\alpha^2)}}
                    \right]\label{majo3}.
\end{eqnarray}
Note that the redefined parameters $p/k$
and $q/k$ are functions of $\alpha$, $\beta$ and $\phi_d$, while the mass
eigenvalues $m_i$, depend on $k$ as well.

\begin{figure}[!h]
\begin{center}
\includegraphics[scale=0.7]{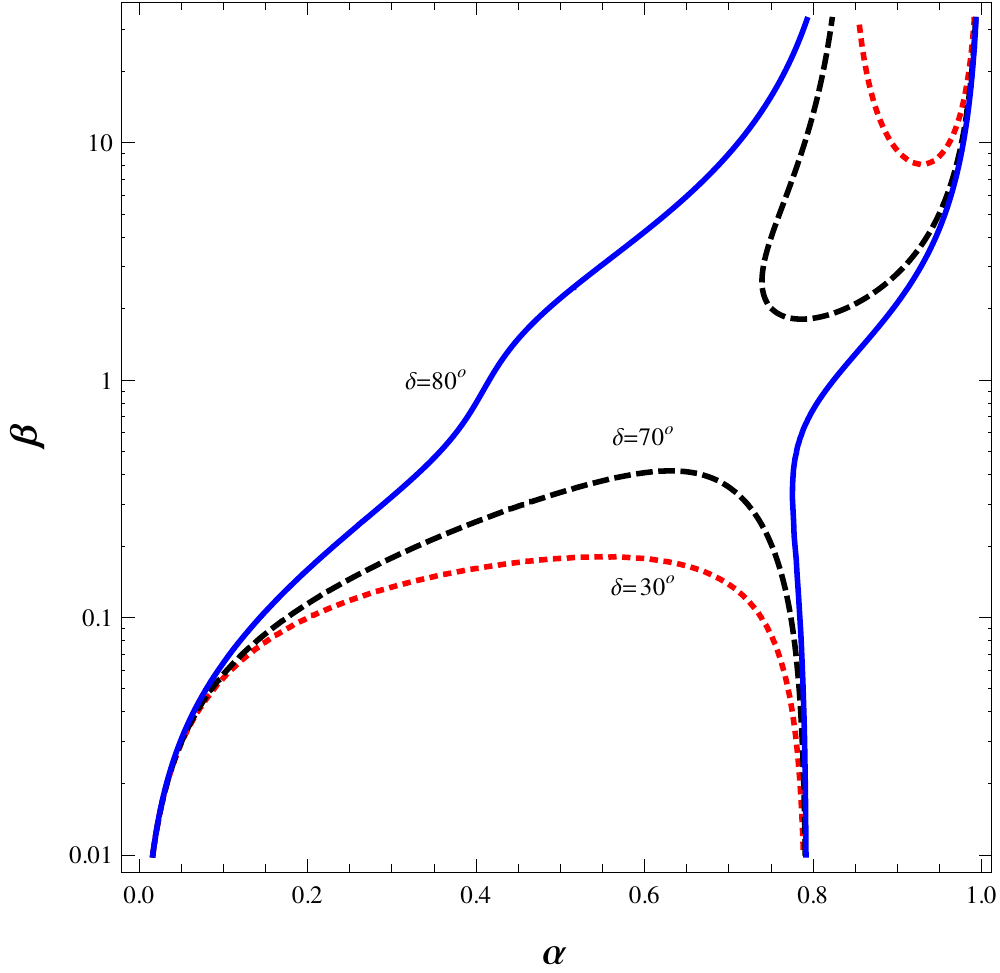}
\includegraphics[scale=0.7]{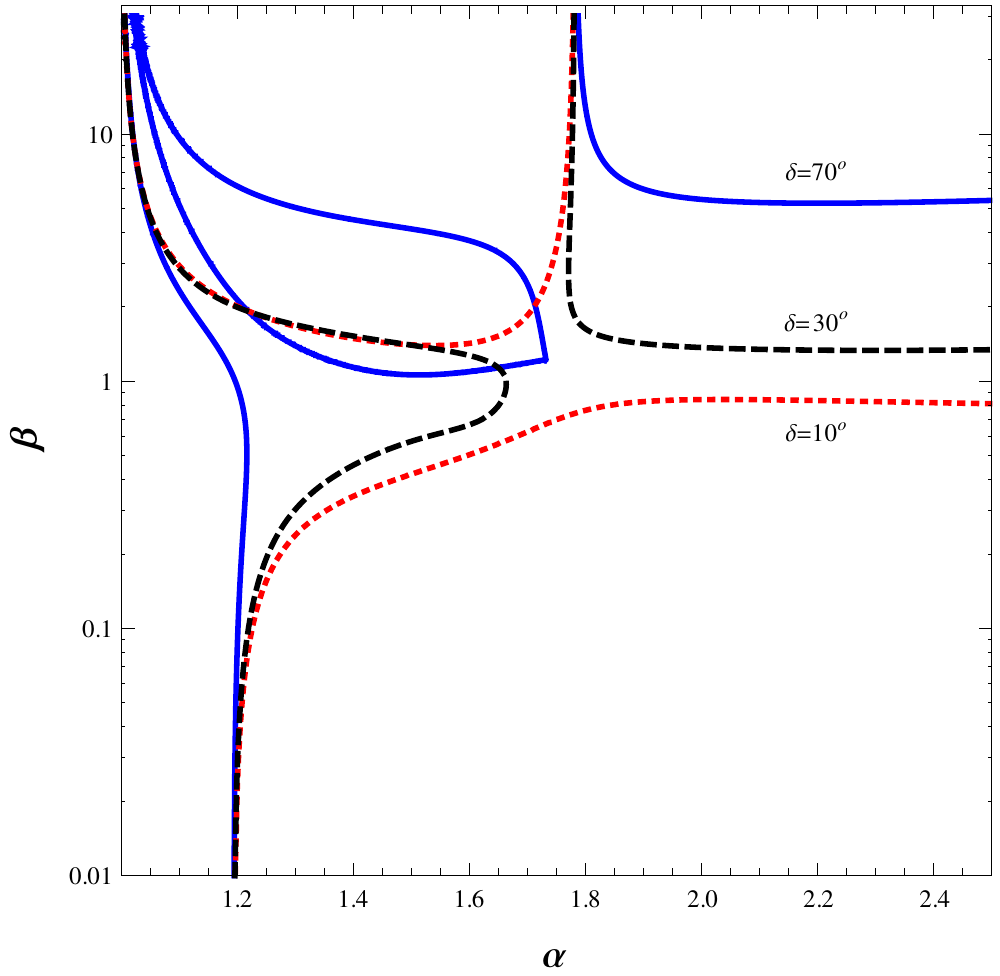}
\caption{{\small Contour plots for $r=0.03$ are shown in the $\alpha-\beta$ plane.
          Here in the left panel (with $\alpha<1$, case A) red (dotted),
          black (dashed) and blue (continuous) lines represent
         $\delta=30^\circ,70^\circ$ and $80^\circ$ respectively. Similar contours
are present for $|\pi - \delta|$ values of the CP violating Dirac phase.
In the right panel (with $\alpha>1$, case B) red (dotted),
          black (dashed) and blue (continuous) lines represent
         $\delta=10^\circ,30^\circ$ and $70^\circ$ respectively. }}
\label{rl1}
\end{center}
\end{figure}

The parameters $\alpha, \beta$ and $\phi_d$ can now be constrained 
by the neutrino oscillation data. To have a 
more concrete discussion, we consider the ratio, $r$, defined by
$r = \frac{\Delta m^2_{\odot}}{\vert \Delta m^2_{atm} \vert}$, 
with $\Delta m^2_{\odot} \equiv \Delta m^2_{21} = m^2_2 - m^2_1$ and  
$\vert \Delta m^2_{atm} \vert \equiv \Delta m^2_{31} = m^3_3 - m^2_1$ 
considering normal hierarchy. Following \cite{Forero:2014bxa}, the best
fit values of $\Delta m^2_{\odot} = 7.6\times 10^{-5}\hspace{.1cm} \eVq$
and $|\Delta m^2_{atm}| = 2.48\times 10^{-3}\hspace{.1cm} \eVq$ are used 
for our analysis. Using Eqs.(\ref{m1}-\ref{m3}), we have an expression 
for $r$ as, 
\begin{eqnarray}
 r&=&\frac{\pm(1-\alpha^2)}{4\alpha }\frac{k}{p}\label{rex}
     \left[
               1+\beta^2-2\beta\cos\phi_d
               -
\left( \frac{\alpha}{\pm(1-\alpha^2)}  - \frac{p}{k} \right)^2
                   - \left(\frac{q}{k}\right)^2\
                \right].
\end{eqnarray}
Here also, `$+$' corresponds to case A ($i.e.$ with $\alpha<1$) and `$-$' is for 
case B ($i.e.,$ when $\alpha>1$). 
Interestingly we note that $r$ depends on $\alpha, \beta$ and $\phi_d$. 
Therefore using this expression of $r$, we 
can now have a contour plot for $r = 0.03$ \cite{Agashe:2014kda} in terms of
$\alpha$ and $\beta$ for specific choices of $\delta$ as 
we can replace the $\phi_d$ dependence in terms of $\alpha$ and $\delta$ through 
Eq.(\ref{delta-phid}).
For $\alpha < 1$, this is shown in Fig.\ref{rl1}, left panel and a 
similar plot is made for $\alpha > 1 $ in the right panel. Although we 
argue that it is more natural to consider $\beta$ to be less than one, in
this plot we allow larger values of $\beta$ as a completeness. With this, 
for $\alpha<1$ (case A) we see the appearance of two separate contours of 
$r = 0.03$ with $\delta=30^{\circ}$, one is for $\beta <1$
and the other corresponds to $\beta > 1$. Similar plots are obtained 
for $\delta=70^{\circ}$ as well. However these isolated contours become a 
connected one once the value of $\delta$ increases, e.g. at $\delta =
80^{\circ}$, it is shown in Fig.2, left panel. A similar pattern follows in 
case of $\alpha >1$ case. Below we discuss the predictions of our model for
case A (with $\alpha < 1$) and case B ($\alpha > 1$) separately.

%

\subsection{Results for Case A}
\begin{figure}[!h]
\begin{center}
\includegraphics[scale=0.7]{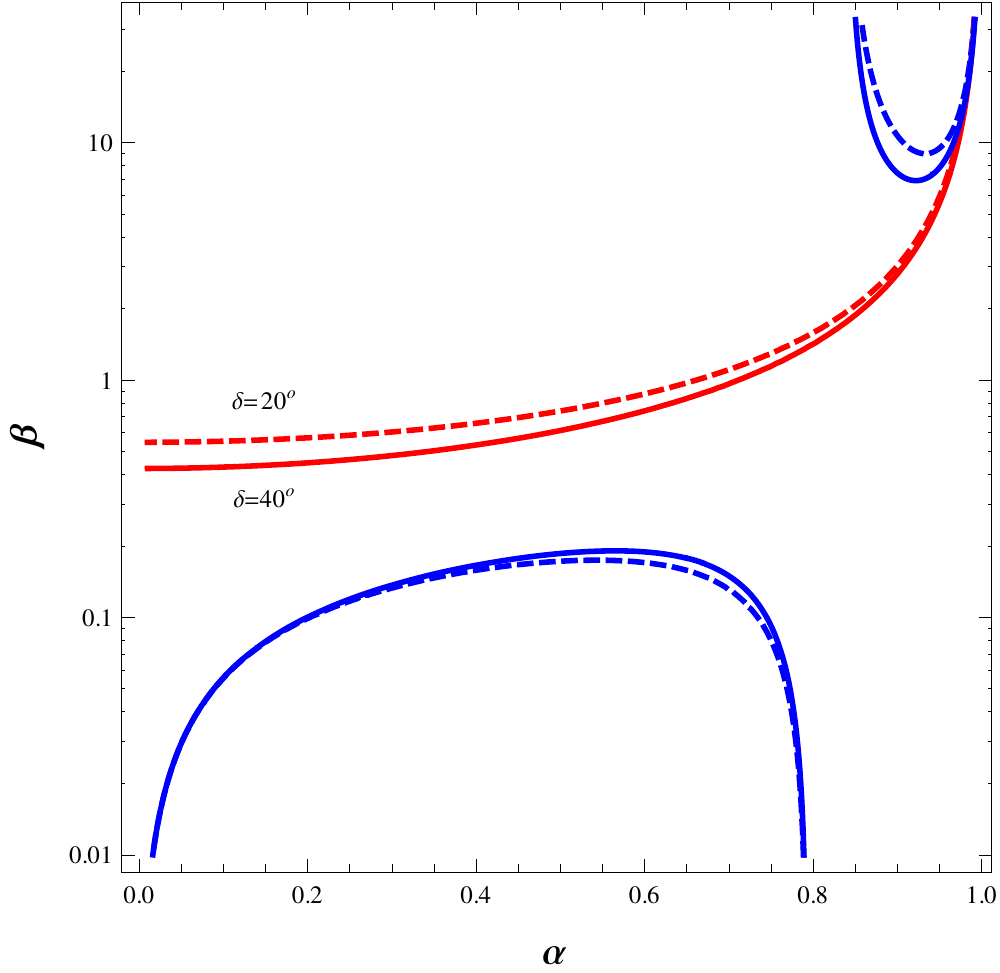}
\includegraphics[scale=0.7]{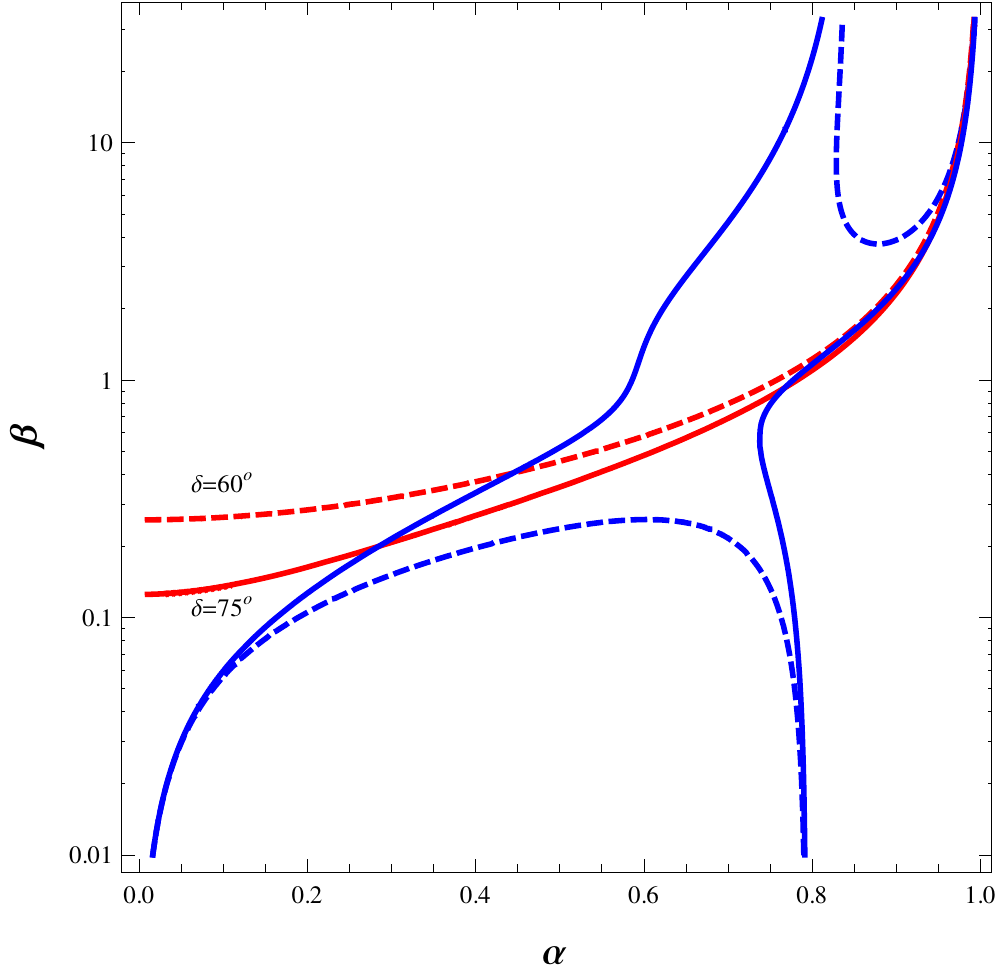}
\caption{{\small Contour plots for both $\sin^2\theta_{13}=0.0234$ and $r=0.03$
         in the $\alpha-\beta$
         plane for various choices of $\delta$. In the left panel,
         dotted and continuous  lines
         represent $\delta=20^\circ$ and $40^\circ)$ respectively.
         In the right panel, dashed and continuous lines
         represent contour plots for
         $\delta=60^\circ$ and $75^\circ$ respectively.}}
\label{2040}
\end{center}
\end{figure}

Note that we need to satisfy both the $\sin^2\theta_{13}$ as well as the 
value of $r$ obtained from the neutrino oscillation experiments. For this reason, if
we consider the two contour plots (one for $r =0.03$ and the other for 
$\sin^2 \theta_{13} =0.0234$) together, then their intersection (denoted by
($\alpha$, $\beta$)) should indicate a simultaneous satisfaction of these 
experimental data for a specific choice of $\delta$.  This is exercised in 
Fig.{\ref{2040}}. In the left panel of Fig.\ref{2040}, contour plots of $r$ and
$\sin^2\theta_{13}$ are drawn in terms of $\alpha$ and $\beta$ for two 
choices of $\delta$= $20^{\circ}$ and $40^{\circ}$. 
We find that there is no such solution for $(\alpha, \beta)$ which satisfy
both $r$ and $\sin^2{\theta_{13}}$ with $\alpha , \beta \lesssim 1$ in these cases.
However there exists solution for $\alpha$ very close to one with a pretty large
value of $\beta$ as mentioned in Table 3. This solution as we expect is not a natural one, 
not only for a large value of $\beta$, but also for its very fine tuned situation. 
Note that $\alpha$ requires to be sufficiently close (and hence finely tuned) to one in 
this case. This situation can be understood from the fact that $\beta$ being
quite large ($\gg 1$), value of $\alpha$ has to be adjusted enough (see the
involvement of the expression $\alpha/(1-\alpha^2)$ in Eq.(\ref{rex})) so as to compete with 
the $\beta$ dependent terms to get $r \sim 0.03$. Similarly variation of
$\sin^2{\theta_{13}}$ is very sharp with respect to $\alpha$ (when close
to 1) for large $\beta$. For example, a small change in $\alpha$ values
($\sim 1 \%$) would induce a change in $\sin^2{\theta_{13}}$  by an amount 
of $15\%$ near the intersection region. 

However the situation changes dramatically as we proceed for higher values
of $\delta$ as can be seen from Fig.{\ref{2040}}, right panel. This figure is
for two choices of $\delta$= $ 60^\circ$ and 75$^\circ$. We observe that with
the increase of $\delta$, the upper contour for $r$ is extended toward 
downward direction and the lower one is pushed up, thereby providing a 
greater chance to have an intersection with the $\sin^2\theta_{13}$ 
contour. We also note that the portion of 
$\sin^2{\theta_{13}}$ contour for $\alpha < 1$ prefers a region with 
relatively small value of $\beta  (<1)$ as well. However a typical solution with
both $\alpha$  and $\beta < 1$ appears when $\delta$ is closer to 
$75^\circ$. With this $\delta$, we could see the lower and upper contours
open up to form a connected one and we can have a solution for $(\alpha,
\beta)\equiv(0.29, 0.2)$. In this case, there is one more intersection
between the $r$ and $\sin^2{\theta_{13}}$ contours with $\alpha, \beta<1$
as given by (0.77, 0.93). When $\delta$ approaches $ 80^\circ$ and up 
(till $\pi/2$) we have have solutions with $\alpha, \beta < 1$. 

\begin{table}[h]\centering
  \begin{tabular}{|c|c|c|c|c|}
    \hline
   $\delta$ & $\alpha$ &$\beta$&$\sum m_i$(eV)\\
    \hline\hline
     $20^\circ (160^\circ,200^\circ,340^\circ)$ & 0.99 &28.26&0.0714\\
   \hline
    $40^\circ(140^\circ,220^\circ,320^\circ)$ & 0.99 &20.94&0.0709\\
    \hline
   $60^\circ(120^\circ,240^\circ,300^\circ)$ & 0.98 &11.16&0.0701\\
    \hline
     $75^\circ(105^\circ,255^\circ,285^\circ)$ & \begin{tabular}{c}
                                 0.94 \\
                                 0.77\\
                                 0.29
                           \end{tabular}      &\begin{tabular}{c}
                                                 3.70 \\
                                                 0.93\\
                                                 0.20
                                               \end{tabular} 
                                                &\begin{tabular}{c}
                                                 0.0691\\
                                                 0.0734\\
                                                 0.1333
                                               \end{tabular}\\
    \hline
   $80^\circ(100^\circ,260^\circ,280^\circ)$ & 0.16 &0.11&0.1835\\
   \hline
   $82^\circ(98^\circ,262^\circ,278^\circ)$ & 0.12 &0.09&0.2137\\
    \hline
   $85^\circ(95^\circ,265^\circ,275^\circ)$ & 0.07 &0.05&0.2827\\ 
    \hline
     \end{tabular}
     \caption{ \label{tab3} {\small$\alpha, \beta$ values at the intersection points of
     the $r$ and $\sin^2 \theta_{13}$ contour plots are provided corresponding to different 
     $\delta$ values. The sum of the light neutrino masses are also indicated in each case.}}
\end{table}

We have scanned the entire range of $\delta$, from $0$ to $2\pi$ and listed
our findings in Table \ref{tab3}. For the $\delta$ values, we denote inside 
the first bracket those values of $\delta$, for which the same set of solution 
points ($\alpha, \beta$) are obtained. This is due to the fact that corresponding 
to a $r$ or $\sin^2{\theta_{13}}$ contour plot for a typical $\delta$ between 
0 and $2\pi$, the same plot is also obtained for other $|\pi - \delta|$ values. 
Accepting the solutions for which $\alpha, \beta < 1$ (i.e. those are not fine tuned 
with large $\beta$), we find that the our setup then predicts an acceptable 
range of CP violating Dirac phase $\delta$ to be between  $72^{\circ}-82^{\circ}$, 
while the first quadrant is considered. For the whole range of $\delta$ 
between 0 and $2 \pi$, the allowed range therefore covers $72^{\circ}-82^{\circ}$, 
$98^{\circ}-108^{\circ}$, $252^{\circ}-262^{\circ}$, $278^{\circ}-288^{\circ}$. 
Note that $\delta$ between $83^{\circ}$ and $90^{\circ}$ (similarly regions of $\delta$ in 
other quadrants also) is ruled out from the constraints on the sum of the 
light neutrino mass mentioned in Table\ref{tab3}. We will discuss about
it shortly. Also the values of $\delta$s like 0, $\pi$, $2\pi$ are disallowed in our 
setup as they would not produce any CP violation which is the starting point 
of our scenario. Again $\delta = \pi/2, 3\pi/2$ are not favored as we have not 
obtained any solution of $\alpha, \beta$ that satisfied both $r$ and 
$\sin^2{\theta_{13}}$. The same is true for the case with $\alpha > 1$. 

We will now proceed to discuss the prediction of the model for the light
neutrino masses and other relevant quantities in terms of the parameters
involved in the set-up. For this, from now onward, we stick to the choice 
of $\delta = 80^{\circ} (\equiv 100^{\circ}, 260^{\circ}, 280^{\circ})$
as a reference value for the Dirac CP violating phase. 
\begin{figure}[!h]
\begin{center}
\includegraphics[scale=0.7]{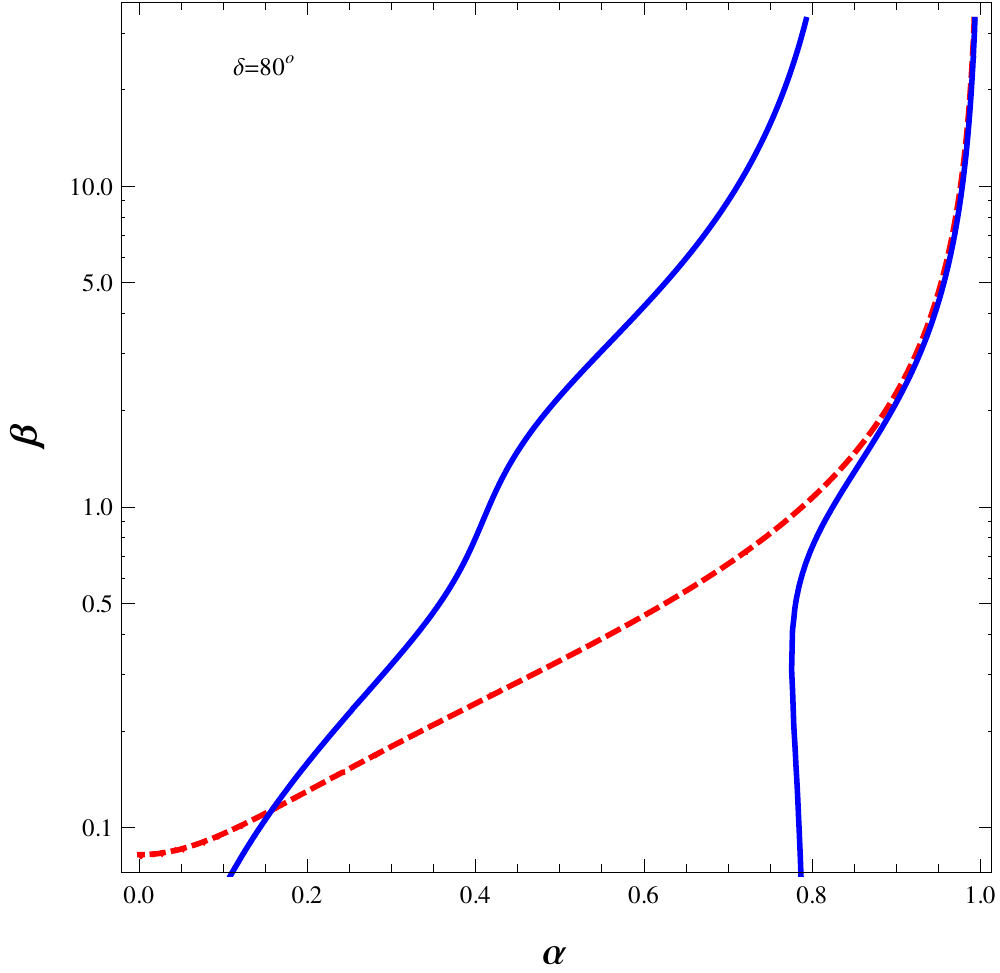}
\includegraphics[scale=0.7]{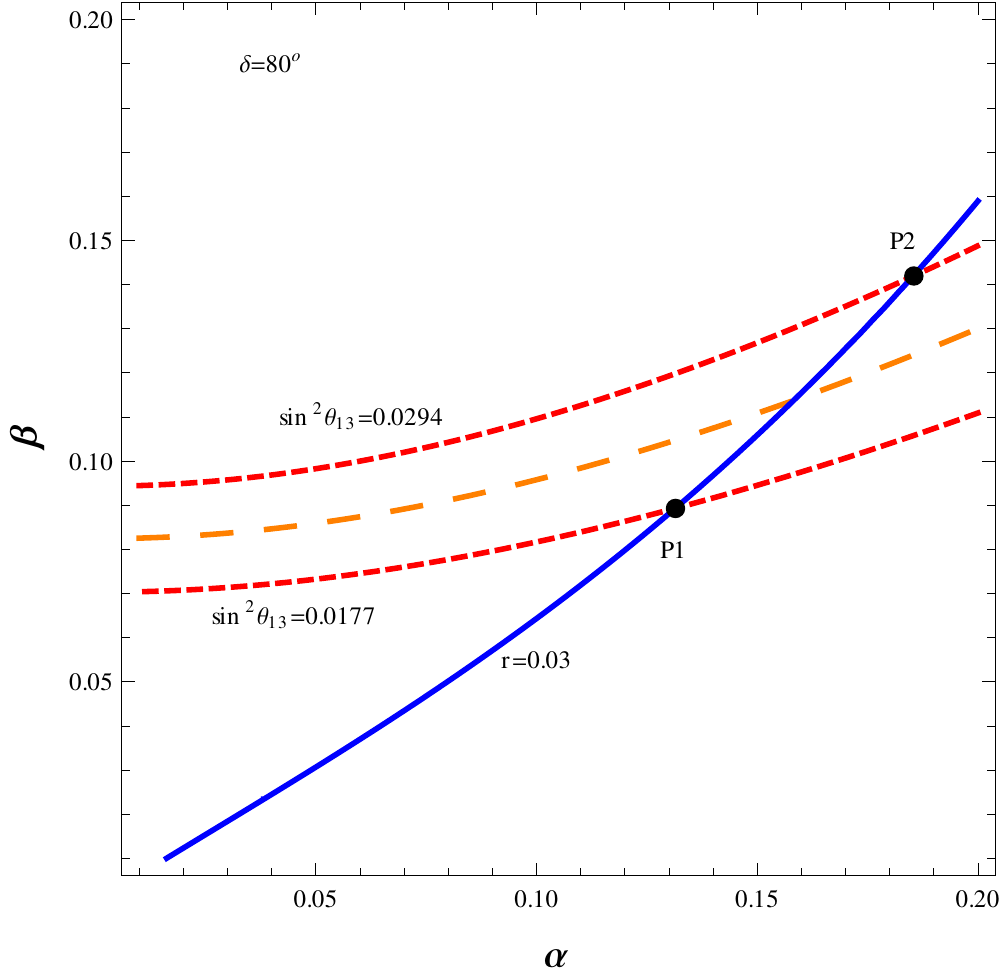}
\caption{{\small Left panel contains contour plots for best-fit
         values of $r$ and $\sin^2\theta_{13}$ for $\delta=80^\circ$ in
         $\alpha$-$\beta$ plane. Here red dashed and blue continuous 
         lines represent contours for $\sin^2\theta_{13}$ and $r$ 
         respectively. The right panel is for best fit value of $r$
         (blue continuous line) and 3$\sigma$ range of $\sin^2\theta_{13}$
         (denoted by two red dotted lines).
         }}
\label{80}
\end{center}
\end{figure}
The $r$ and $\sin^2\theta_{13}$ contours for this particular
$\delta$ is shown separately in Fig.(\ref{80}), left panel. In Fig.(\ref{80}) right
panel we put the $\sin^2{\theta_{13}}$ contours corresponding 
to the upper and lower values (detonated by red dotted lines)  those are 
allowed by the 3$\sigma$ range of $\sin^2{\theta_{13}}$ .
Only a section of  $r$ contour is also incorporated which encompasses
the ($\alpha,\beta$) solution points. This plot provides a range for 
$(\alpha, \beta)$ once the 3$\sigma$ patch of $\sin^2{\theta_{13}}$ are 
considered. It starts from a set of values $(\sim 0.13,0.09)$ (can be 
called a reference point P1) upto $(\sim 0.18,0.14)$ (another reference
point P2). Note that there is always a one-to-one correspondence between
the values of $\alpha$ and $\beta$, which falls on the line of $r$ contour. 

We have already noted that in the expression for $r$, parameters $\alpha,
\beta$ and $\phi_d$ are present. Once we choose a specific $\delta$, 
automatically it boils down to find $\alpha$ and $\beta$ from 
Eq.(\ref{rex}). Although $r$ is the ratio between $\Delta m^2_{\odot}$ and 
$|\Delta m^2_{atm}|$, we must also satisfy the mass-squared differences
$\Delta m^2_{\odot}$ as well as $|\Delta m^2_{atm}|$ independently. For that
we need to determine the value of the $k$ parameter itself apart from its
involvement in the ratio $\beta = \vert d \vert /k$ as evident from 
Eq.(\ref{m1})-(\ref{m3}). For this purpose, with $\delta = 80^{\circ}$ 
while moving from P1 to P2 along the $r$ contour in the right panel of
Fig. \ref{80}, we find the values of $\alpha$ and correspondingly 
$\beta$ which produce $r = 0.03$. Now using these values of ($\alpha, \beta$),
we can evaluate the values of $k$ for each such set which satisfies 
$\Delta m^2_{\odot}=7.6\times10^{-5}\eVq$. To obtain these values of $k$
corresponding to ($\alpha, \beta$) set, we employ Eqs.(\ref{m1}-\ref{m2}). 
\begin{figure}[!h]
\begin{center}
\includegraphics[scale=0.7]{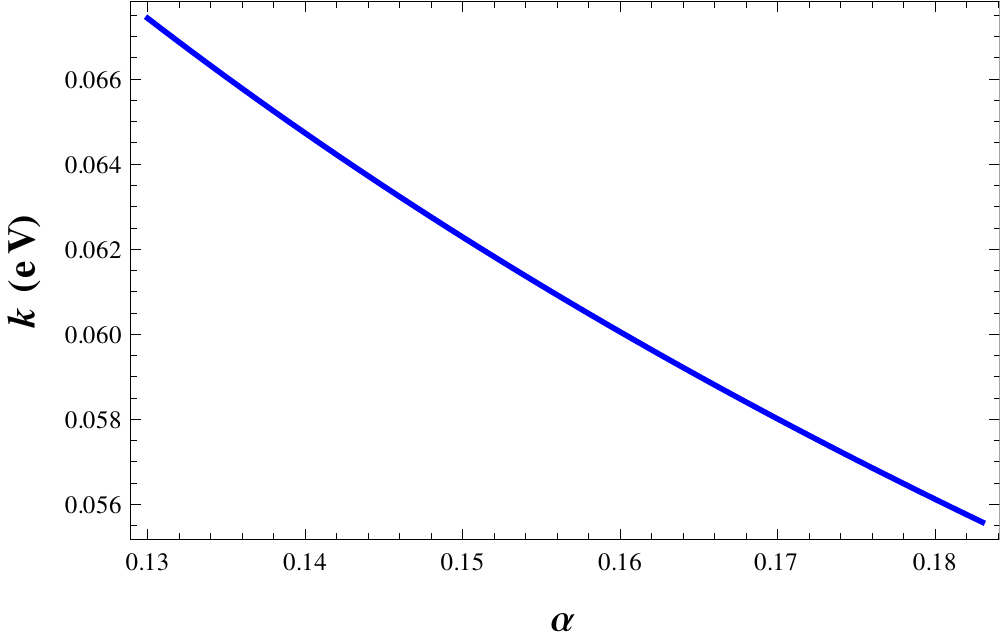}
\includegraphics[scale=0.7]{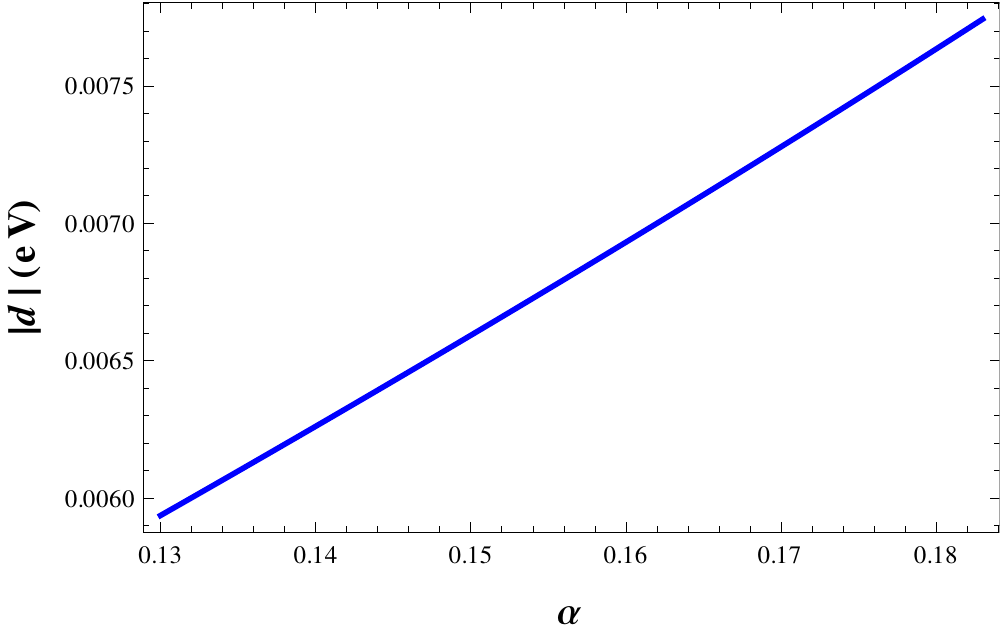}
\caption{{\small $k$ vs. $\alpha$ (left-panel) and $|d|$ vs. $\alpha$
                  (right-panel) for $\delta=80^\circ(\equiv 100^\circ,260^\circ,280^\circ)$. }}
\label{kda}
\end{center}
\end{figure}
The result is reflected in left panel of Fig.{\ref{kda}}, where we plot the
required value of $k$ in terms of its variation with $\alpha$. In producing
the plot, only a narrow range of $\alpha$ is considered which corresponds to
the 3$\sigma$ variation of $\sin^2\theta_{13}$ as obtained from Fig.(\ref{80}),
right panel ($i.e.$ from P1 to P2). Although we plot it against $\alpha$, 
each value of $\alpha$ is therefore accompanied by a unique value of $\beta$,
as we just explain. Once the variation of $k$ in terms of $\alpha$ is known,
we plot the variation of $\vert d \vert$ ($\beta$= $\beta k$) with $\alpha$
in Fig.{\ref{kda}}, right panel. Having the correlation between $\alpha$ 
and other parameters like $\beta, k$ for a specific choice of $\delta$ is known, 
we are able to plot the individual light neutrino masses using Eqs.(\ref{m1}-\ref{m3}).
This is done in Fig.{\ref{misf}}. The light neutrino masses satisfies normal mass 
hierarchy. We also incorporate the sum of light neutrino masses ($\Sigma m_i $)
to check its consistency with the cosmological limit set by Planck, 
$\Sigma m_i < 0.23$ eV \cite{Ade:2013zuv}. In this particular case with 
$\delta = 80^{\circ}$ (also for $\delta = 100^{\circ},260^{\circ},280^{\circ}$),
this limit is satisfied for the allowed range of $\alpha$, it turns out that 
$\delta=83^\circ$ and $97^\circ$ (and similarly for $263^\circ-277^\circ$) do not satisfy 
it as indicated in Table.\ref{tab3}. 
\begin{figure}[!h]
\begin{center}
\includegraphics[scale=0.7]{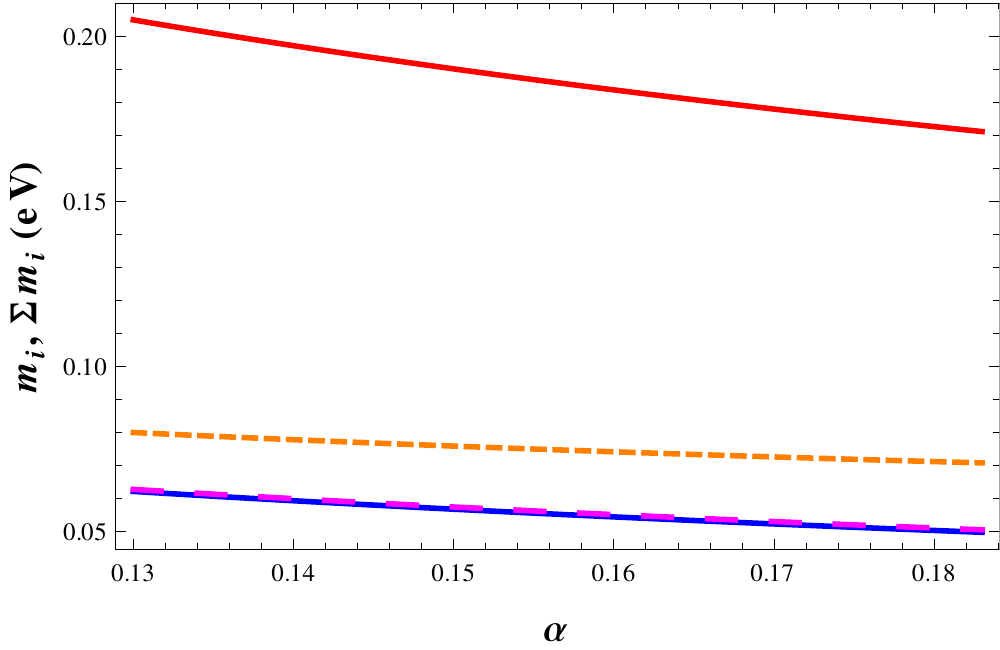}
\caption{{\small Light neutrino masses: $m_1$ (blue continuous line), $m_2$ (magenta large
                 dashed), $m_3$ (orange dashed) and
           $\Sigma m_i$ (red continuous line) vs $\alpha$ for
           $\delta=80^\circ(\equiv 100^\circ,260^\circ,280^\circ)$.}}
\label{misf}
\end{center}
\end{figure}

\begin{figure}[!h]
\begin{center}
\includegraphics[scale=0.7]{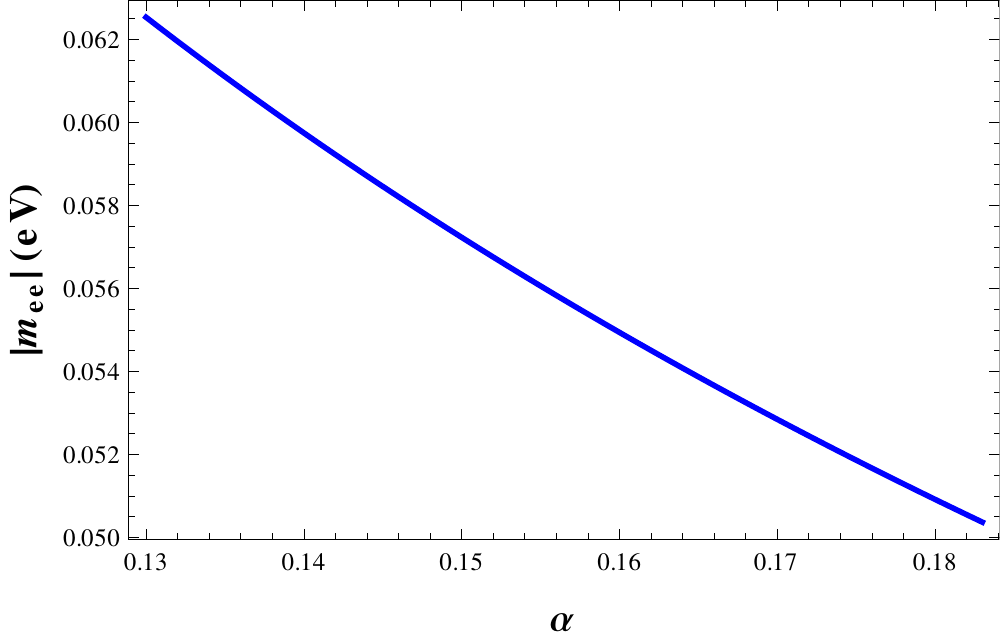}
\includegraphics[scale=0.7]{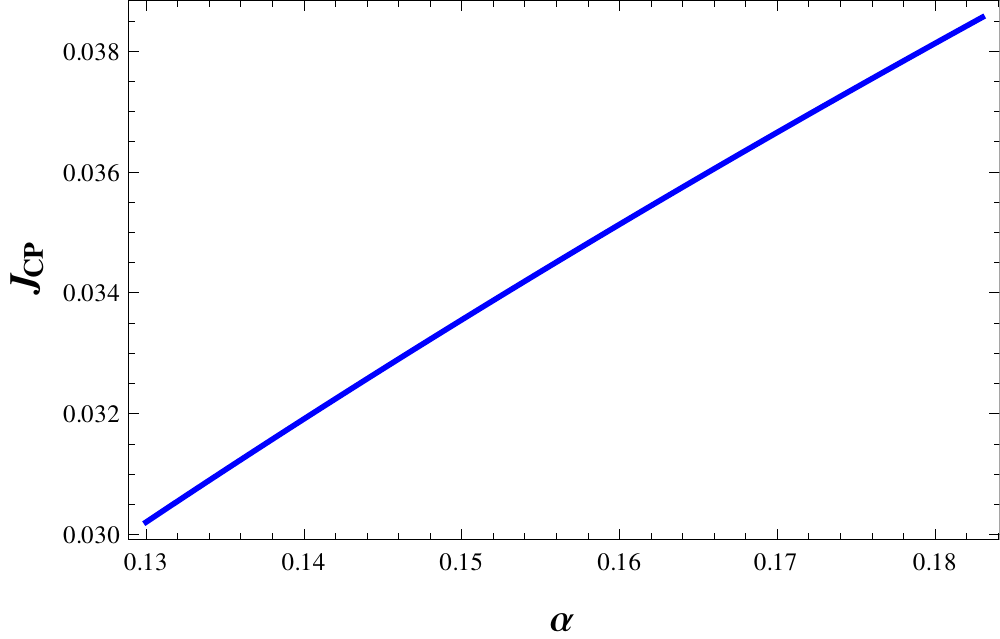}
\caption{{\small Effective neutrino mass parameter (left panel) and
                 Jarlskog invariant (right panel) $\alpha$ for 
                 $\delta=80^\circ(100^\circ,260^\circ,280^\circ)$.}}
\label{meef}
\end{center}
\end{figure}

Now by using Eqs.(\ref{majo1}-\ref{majo3}), we estimate the Majorana
phases\footnote{The source of these phases are the phase $\phi_d$ only.} 
$\alpha_{21}$ and $\alpha_{31}$ for $\delta = 80^{\circ}$, which appears 
in the effective neutrino mass parameter $\vert m_{ee} \vert$. 
$\left|m_{ee}\right|$ appears in evaluating the neutrinoles double beta decay and is given 
by\cite{Agashe:2014kda}, 
\begin{eqnarray}
\left|m_{ee}\right|=\left|m_1^2c_{12}^2c_{13}^2+
m_2^2s_{12}^2c_{13}^2e^{i\alpha_{21}}+
m_3^2s_{13}^2e^{i(\alpha_{31}-2\delta)}\right|. 
\end{eqnarray}
In Fig.{\ref{meef}}, we plot the prediction of $\vert m_{ee}\vert$ 
against $\alpha$ within its narrow range satisfying 3$\sigma$ range of
$\sin^2\theta_{13}$ with $\delta=80^{\circ}$. Here we obtain $0.050 \leq
\vert m_{ee}\vert \leq 0.062$. This could be probed in
future generation experiments providing a testable platform of the model
itself.

It is known that presence of nonzero Dirac CP phase can trigger CP violation
in neutrino oscillation at low energy. In standard parametrization, the 
magnitude of this CP violation can be estimated \cite{Agashe:2014kda} through
\begin{eqnarray}
 J_{CP}&=&{\rm Im}[U_{\mu 3}U_{e 3}^*U_{e 2}U_{\mu 2}^*]\nonumber\\
       &=&\frac{1}{8}\cos\theta_{13}\sin2\theta_{12}\sin2\theta_{23}
          \sin2\theta_{13}\sin\delta. \label{jcp1}
 \end{eqnarray}
As in our model, the unique source of $\delta$ is the CP violating phase 
$\alpha_{\sm}$ in $S$, it is interesting to see the prediction of our model
towards $J_{CP}$. Using the expression of $J_{CP}$ in Eq. (\ref{jcp1})
along with Eqs.(\ref{ang}) and (\ref{ang22}) we estimate $J_{CP}$ in our 
model as shown in Fig.\ref{meef}, right panel with $\delta=80^\circ$.
Here also we include only that range of $\alpha$ which provides solutions 
corresponding to 3$\sigma$ allowed range of $\sin^2\theta_{13}$. However 
we scanned the entire range of $\alpha$ where the solutions exists for all 
allowed values of $\delta$ and find that $J_{CP}$ in our model 
is predicted to be $0.03<|J_{CP}|<0.04$. This can be measured in future 
neutrino experiments.

\subsection{ Results for Case B}\label{sec:ag1}
\begin{figure}[!h]
\begin{center}
\includegraphics[scale=0.7]{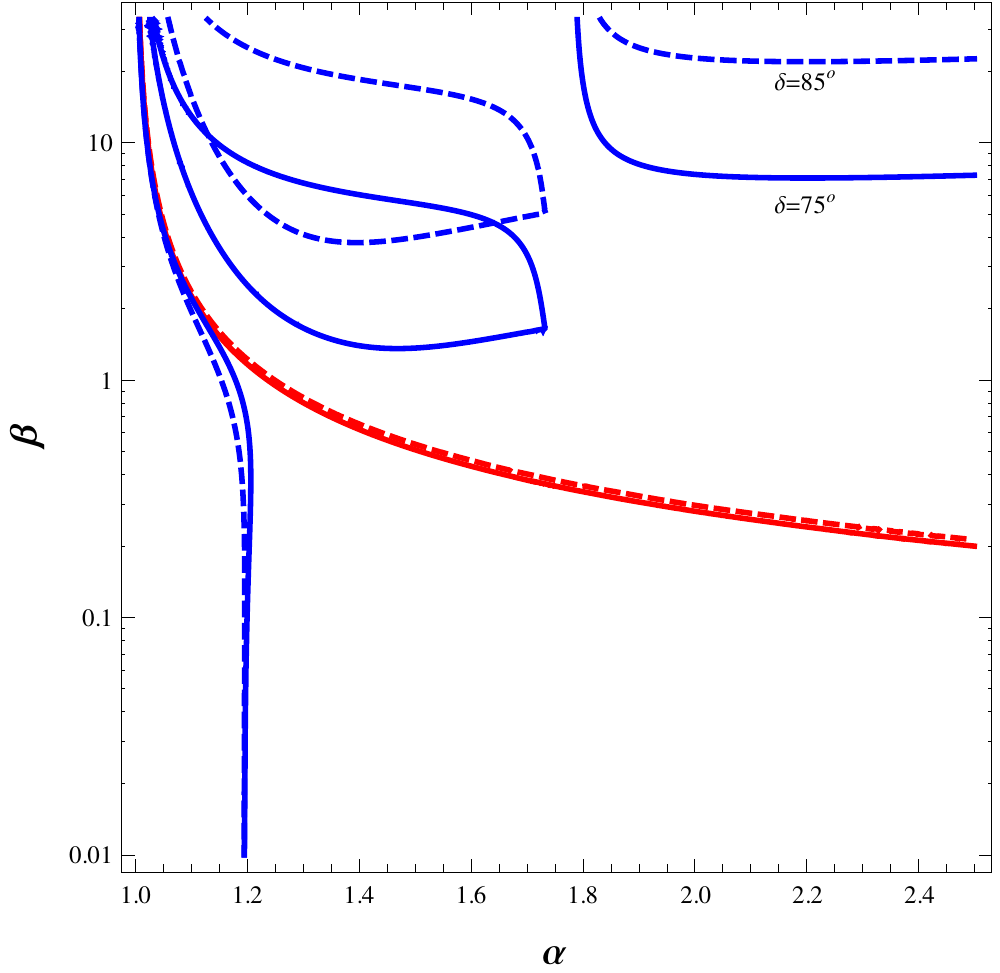}
\includegraphics[scale=0.7]{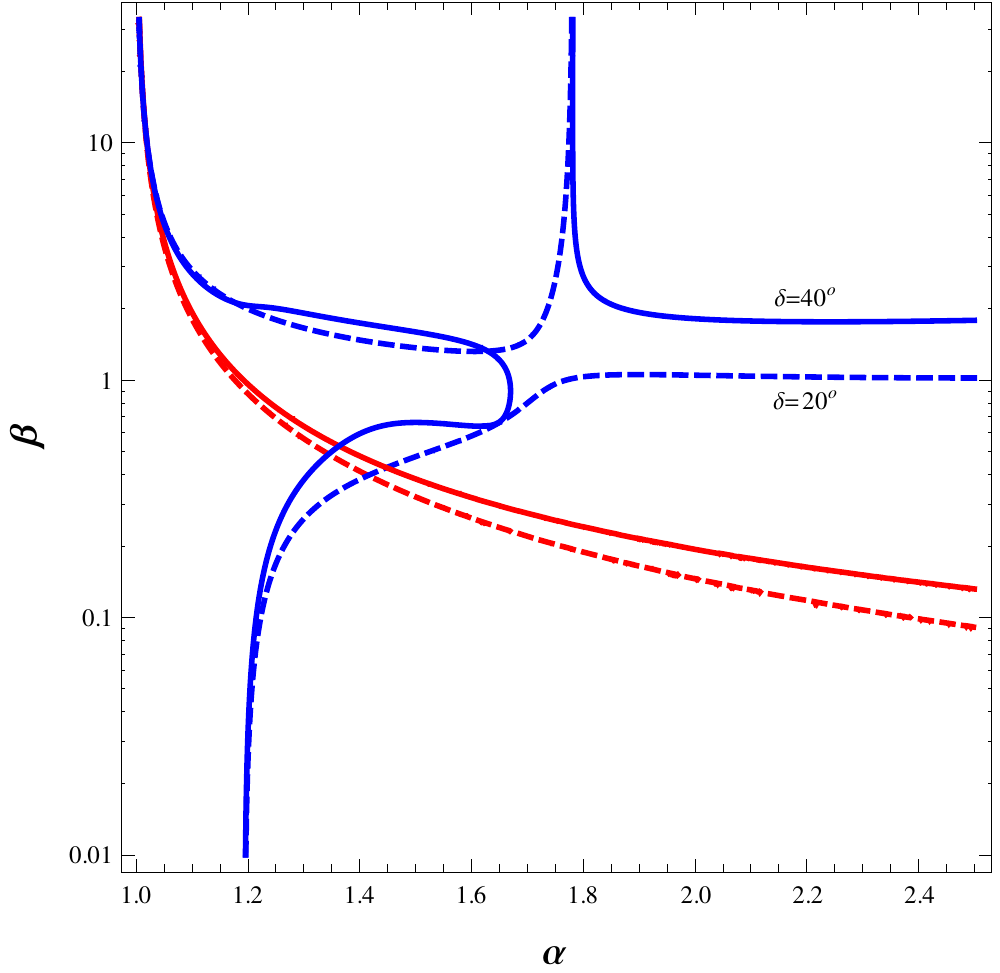}
\caption{{\small Contour plots $r$ and $\sin^2\theta_{13}$ when $\alpha>1$.
         In the left panel dashed (continuous)  line 
         represents $\delta=85^\circ (75^\circ)$.
         Where as, in the right panel,continuous (dashed)   lines
         represents contour plots for 
         $\delta=30^\circ$ ($60^\circ$).}}
\label{85-75}
\end{center}
\end{figure}

\begin{figure}[h]
\begin{center}
\includegraphics[scale=0.7]{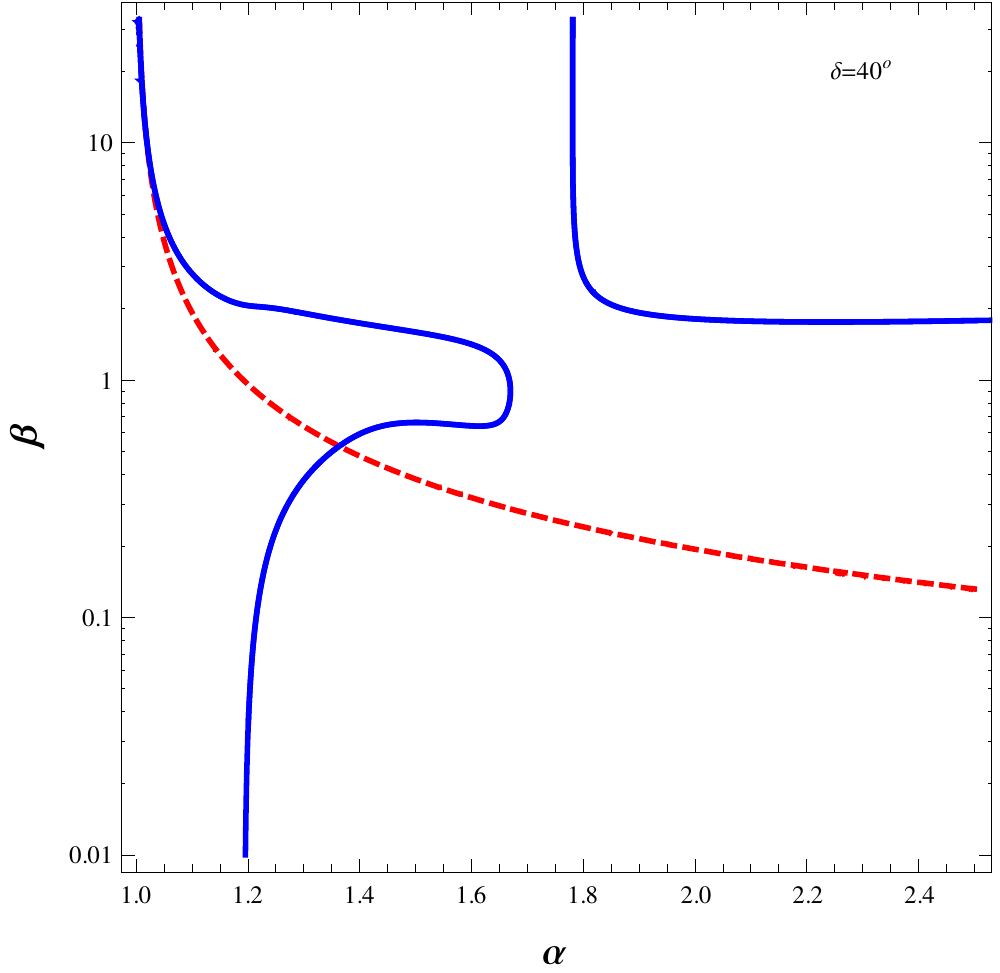}
\includegraphics[scale=0.7]{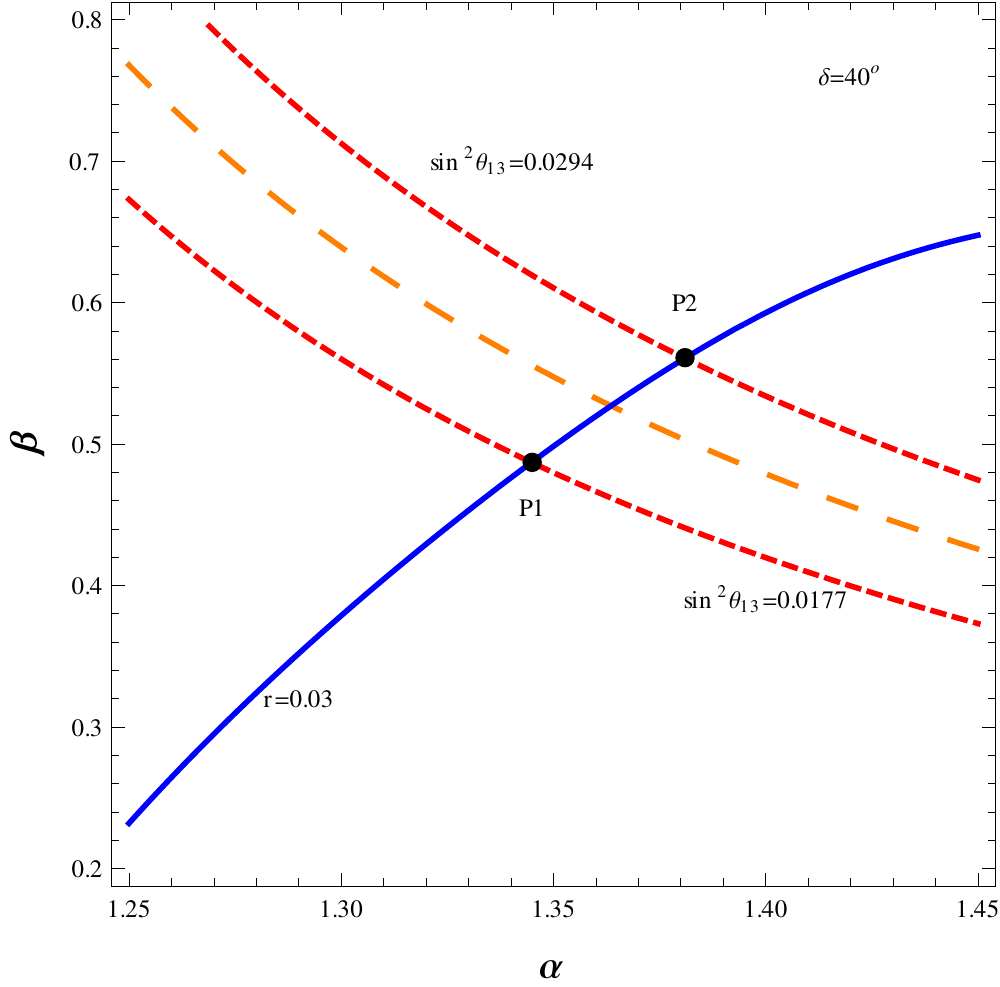}
\caption{{\small  Contour plots of $r =0.03$ and $\sin^2\theta_{13} = 0.0234$ together
           in the $\alpha - \beta$ plane. In the right panel, the intersection region
           is elaborated where 3$\sigma$ regions of
$\sin^2 \theta_{13}$ are depicted.}}
\label{ag1p1}
\end{center}
\end{figure}

Similar to case A, we consider here the expression of $r$ for $\alpha > 1$
from Eq.(\ref{rex}) to draw the contour plot for $r=0.03$ in the $\alpha - \beta$ 
plane as shown in Fig.(\ref{85-75}) while $\delta$ is fixed at different values.
In the same plot we include the $\sin^2\theta_{13}=0.0234$ contour as well to find
the set of parameters ($\alpha,\beta$) 
corresponding to a fixed $\delta$ which satisfies the best fit values of 
$\sin^2\theta_{13}$ and $r$. Once we restrict $\beta$ to be below one, we 
find the  solutions to exists for $\delta=0^{\circ}-63^{\circ}$, 
($117^{\circ}-180^{\circ}$, $180^{\circ}-243^{\circ}$, $297^{\circ}-360^{\circ}$)
shown in Table \ref{ta1}.  
\begin{table}[h]\centering
  \begin{tabular}{|c|c|c|c|c|}
    \hline
   $\delta$ & $\alpha$ &$\beta$&$\sum m_i$(eV)\\
    \hline\hline
     $10^\circ (170^\circ,190^\circ,350^\circ)$ & 1.43 &0.36&0.0791\\
   \hline
    $30^\circ(150^\circ,210^\circ,330^\circ)$ & 1.39 &0.45&0.0798\\
    \hline
   $40^\circ(140^\circ,220^\circ,320^\circ)$ & 1.36 &0.53&0.0799\\
    \hline
   $50^\circ(130^\circ,230^\circ,310^\circ)$ & 1.32 &0.64&0.0794\\
    \hline
   $60^\circ(120^\circ,240^\circ,300^\circ)$ & 1.26 &0.83&0.0776\\ 
    \hline
    $70^\circ(110^\circ,250^\circ,290^\circ)$ & 1.17 &1.13&0.0739\\ 
    \hline 
    $73^\circ(107^\circ,253^\circ,287^\circ)$ & 1.07 &3.02&0.0696\\ 
    \hline
     \end{tabular}
     \caption{ \label{ta1} {\small Solutions for $\alpha (>1)$ and
     $\beta$ for various $\delta$.}}
\end{table}
For $\delta$'s beyond $63^{\circ}$ (when considered within $\pi/2$), the
solutions exhibit $\beta \gg 1$ implying a fine tuned situation similar to case A.  
Note that $\alpha$ therefore falls in a narrow range $\simeq 1.2-1.4$
in order to satisfy both $\sin^2\theta_{13}=0.0234$ and $r=0.03$ considering all
$\delta$ values. In Fig.(\ref{ag1p1}), left panel, we find the intersection 
is at ($1.36,0.53$) for $\delta=40^{\circ}(\equiv 140^{\circ}, 220^{\circ}, 
320^{\circ})$. Considering this $\delta$ as a reference for discussion, we 
further include the $3\sigma$ range of $\sin^2\theta_{13}$ in Fig.\ref{ag1p1}, 
right panel. We find $\alpha$ to be varied between 
1.35 and 1.39 while $\sin^2\theta_{13}$ changes from the lower to the 
higher value, within 3$\sigma$ limit. Within this range, we predict 
\begin{figure}[h]
\begin{center}
\includegraphics[scale=0.7]{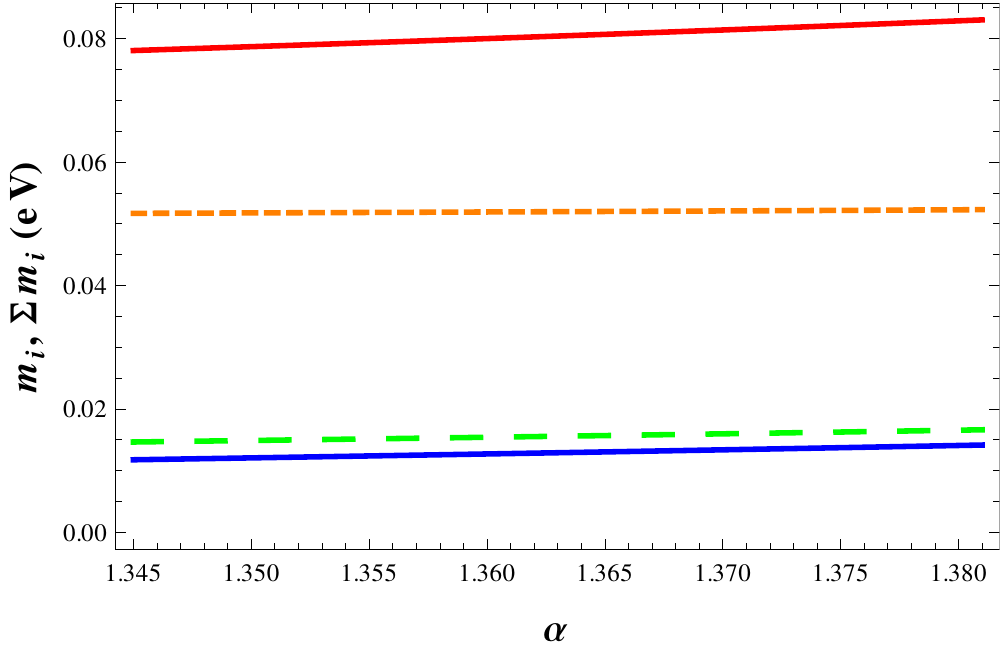}
\caption{{\small Neutrino masses vs $\alpha$ for  $\delta=40^{\circ}( 140^{\circ}, 220^{\circ}, 
320^{\circ})$ when $\alpha>1$}}
\label{mg1}
\end{center}
\end{figure}
individual light neutrino masses and their sum. Here also we find normal
hierarchy for them in Fig.\ref{mg1}. For different $\delta$-values, the $\Sigma m_{i}$
(corresponding to the best fit value of $\sin^2\theta_{13}$) are provided
in Table \ref{ta1}. For showing the prediction of our model in terms of other 
quantities like $|m_{ee}|$ and $J_{CP}$, the Fig.\ref{meefg1},left and right panels are
included. Considering all the $\delta$ values for which $\beta \leq 1$, we find 
$\left| J_{CP} \right|$ to be within $\left| J_{CP}\right| < 0.035$. 
\begin{figure}[!h]
\begin{center}
\includegraphics[scale=0.7]{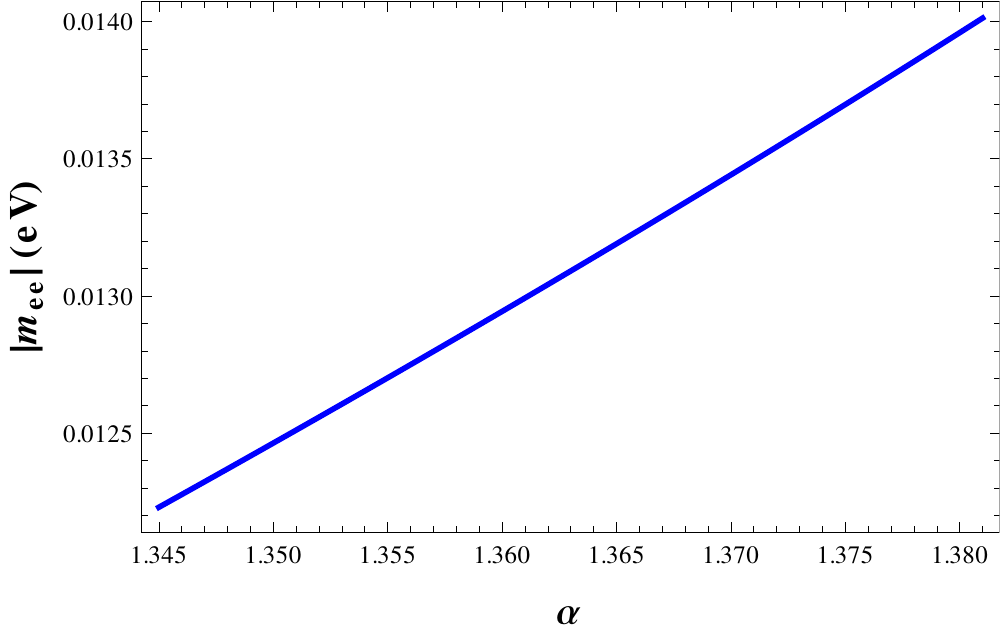}
\includegraphics[scale=0.7]{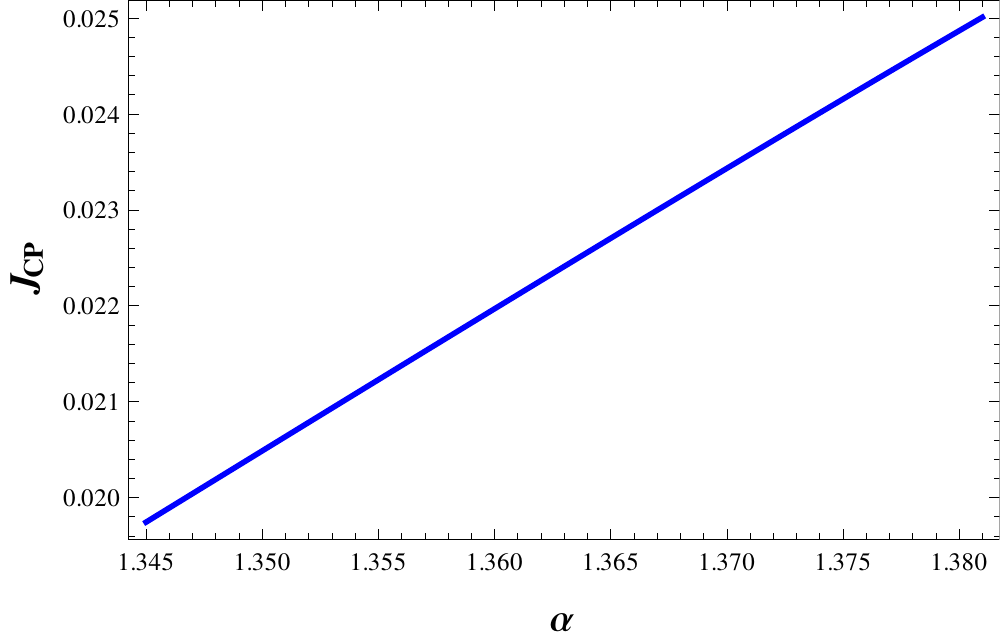}
\caption{{\small Effective neutrino mass parameter (left panel) and
                 Jarlskog invariant (right panel) for 
                 $\delta=40^{\circ}( 140^{\circ}, 220^{\circ}, 
320^{\circ})$ when $\alpha>1$ .}}
\label{meefg1}
\end{center}
\end{figure}

\section{Leptogenesis}\label{sec:lep}

In a general type-II seesaw framework, leptogenesis can be successfully
implemented through the decay of RH neutrinos \cite{Davidson:2008bu} or 
from the decay of the triplet(s) involved \cite{O'Donnell:1993am, 
Ma:1998dx,Hambye:2000ui,Hambye:2005tk,Branco:2011zb} or in a mixed 
scenario where both RH neutrino and the triplet(s) contribute \cite{
AristizabalSierra:2011ab, Akhmedov:2008tb, Hambye:2003ka,
AristizabalSierra:2009ex,Antusch:2007km,AristizabalSierra:2012js}. In the 
present set-up, all the couplings involved in the pure type-I contribution
are real and hence the neutrino Yukawa matrices and the RH neutrino mass
matrices do not include any CP violating phase. Therefore CP asymmetry originated
from the sole contribution of RH neutrinos is absent in our framework. As we
have mentioned earlier, the source of CP violation is only present in the 
triplet contribution and that is through the vev of the $S$ field. However as 
it is known\cite{Hambye:2005tk,Hambye:2012fh}, a single $SU(2)_L$ triplet
does not produce CP-asymmetry. Therefore there are two remaining 
possibilities to generate successful lepton asymmetry \cite{Hambye:2003ka,
Sierra:2014tqa}in the present context; (I) from the decay of the triplet 
where the one loop diagram involves the virtual RH neutrinos and
(II) from the decay of the RH neutrinos where the one loop contribution 
involves the virtual triplet running in the loop. Provided the mass of
the triplet is light compared to all the RH neutrinos ($i.e., M_{\mb}<
M_{\R i}$), we consider option (I). Once the triplet is 
heavier than the RH neutrinos, we explore option (II).

 \begin{figure}[!h]
\begin{center}
\begin{fmffile}{n3}
\begin{fmfgraph*}(110,60)
\fmfleft{i1}
\fmfright{o1,o2}
\fmflabel{$\Delta^{\ast}$}{i1}
\fmflabel{$L_l$}{o1}
\fmflabel{$L_i$}{o2}
\fmf{scalar,tension=1.5}{i1,v1}
\fmf{scalar,tension=.5,label=$H$}{v1,v2}
\fmf{plain,tension=0,label=$N_{\R_k}$}{v2,v3}
\fmf{scalar,tension=.5,label=$H$}{v1,v3}
\fmf{fermion}{v2,o1}
\fmf{fermion}{v3,o2}
\fmffreeze
\fmfshift{70down}{v2}
\fmfshift{71up}{v3}
\end{fmfgraph*}
\end{fmffile}
\vspace{.3cm}
\caption{ One-loop diagram which contributes to the generation of $\epsilon_{\mb}$. }\label{figl}
\end{center}
\end{figure}
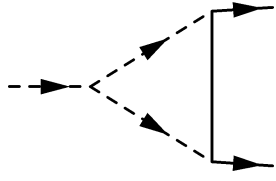

First we consider option (I), $i.e.,$ when $M_{\mb}<M_{\R i}$.
At tree level the scalar triplet can decay either into leptons or into two 
Higgs doublets, followed from the Lagrangian in Eq.(\ref{ld}) and (\ref{sp}).
For $\Delta \longrightarrow L L$, the one loop diagram involves the virtual
RH neutrinos running in the loop as shown in Fig. \ref{figl}. Interference
of the tree level and the one loop results in the asymmetry
parameter \cite{Hambye:2003ka,O'Donnell:1993am,Lazarides:1998iq}
\begin{eqnarray}
 \epsilon_{\mb}&=&2\frac{\Gamma(\Delta^*\longrightarrow L + L )
             -\Gamma(\Delta\longrightarrow \bar{L} + \bar{L})}
             {\Gamma(\Delta^*\longrightarrow L+ L)
             +\Gamma(\Delta\longrightarrow \bar{L} + \bar{L})},\\
         &=&\frac{1}{8\pi}\sum_{k} M_{Rk}\label{ep0}
             \frac{
                 \sum_{il}{\rm Im} [(\hat{Y}^{\ast}_{\D})_{ki}
                  (\hat{Y}^{\ast}_{\D})_{kl}(Y_{\mb})_{il}\eta^{\ast}]
                  }
                  {
                  \sum_{ij}|(Y_{\mb})_{ij}|^2 M^2_{\mb}+|\eta|^2
                  }
                  {\rm log}(1+M^2_{\mb}/M^2_{Rk}).
\end{eqnarray}
Here $i,j$ denote the flavor indices, $\hat{Y}_{\D}=U^{T}_{\R}Y_{\D}$ in the basis where 
RH neutrino mass matrix is diagonal. $Y_{\mb}$, $Y_{\D}$
and expression of $\eta$ can be obtained from Eqs.(\ref{MDMR}),(\ref{YDelta}) and 
(\ref{D1D2}). Masses of RH neutrinos can be expressed as 
\begin{eqnarray}
 M_{\R_1}&=&\frac{v^2y^2}{k}(1+\alpha),\\
  M_{\R_2}&=&\frac{v^2y^2}{k},\\
   M_{\R_3}&=&\left | \frac{v^2y^2}{k}(1-\alpha) \right |.
\label{MR}
 \end{eqnarray}
Therefore, the asymmetry parameter in our model is estimated to be \cite{Hambye:2003ka}
\begin{eqnarray}\label{ep1}
 \epsilon_{\mb} &=& -\frac{M^2_{\mb}}{8\pi v^2}\frac{\alpha^2}{(1-\alpha^2)}
 \frac{
 k\mu\tilde{\omega}^3 v_{\sm}  (x_1-x'_1)\sin\alpha_{\sm}
 }
 {
 \left[3\tilde{\omega}^2\frac{v_{\sm}^2}{\Lambda^2} \left(x_1^2+x_1'^2+2x_1x_1'\cos2\alpha_{\sm}\right) M^2_{\mb}
   +(\mu\tilde{\omega}^2\Lambda)^2\right]
 }.
\end{eqnarray}
Here we denote $\tilde{\omega}=v_{f}/\Lambda$,
where $v_{f}$ is considered to be the common vev of all flavons except $S$-field's 
vev $\langle S \rangle=v_{\sm}e^{i\alpha_{\sm}}$. The associated phase $\alpha_{\sm}$
is the only source of CP-violation here. The total decay width of the triplet $\Delta$ 
(for $\Delta\rightarrow$ two leptons and $\Delta\rightarrow$ two scalar doublets) is
given by 
\begin{eqnarray}\label{tol}
 \Gamma_{\T}&=& \Gamma_{\mb^{\ast} \rightarrow \Ll \Ll}+\Gamma_{\mb^{\ast} \rightarrow \hh \hh}\\
          &=&\frac{M_{\mb}}{8\pi}
      \left[\sum_{ij}\left|(Y_{\mb})_{ij}\right|^2+\frac{|\eta|^2}{M^2_{\mb}}\right].
\end{eqnarray}

Note that there are few parameters in Eq.(\ref{ep1}), $e.g. ~\alpha,k$ which already
contributed in determining the mass and mixing for light neutrinos. Also $\phi_d$
is related with $\alpha_{\sm}$ by Eq.(\ref{phidal}). In the previous section, we
have found solutions for $(\alpha,\beta)$ that satisfy the best fit values of 
$\sin^2\theta_{13}$ and $r$ for a specific choice of $\delta$ (the reference
values $\delta = 80^{\circ}$ for $\alpha < 1$ and $\delta = 40^{\circ}$ for $\alpha > 1$ ). Then
we can find the values of $k$ and $|d|$ corresponding to that specific $\delta$ value. 
These set of $\alpha, \left|d\right|, k$ produce correct order of neutrino mass and mixing
as we have already seen. Here to discuss the CP-asymmetry parameter $\epsilon_{\mb}$,
we therefore choose $\delta=80^{\circ}
(100^{\circ},260^{\circ},280^{\circ})$ for $\alpha<1$ and $\delta=40^{\circ}
(140^{\circ},220^{\circ},320^{\circ})$ for $\alpha>1$. 

\begin{figure}[!h]
\begin{center}
\includegraphics[scale=0.7]{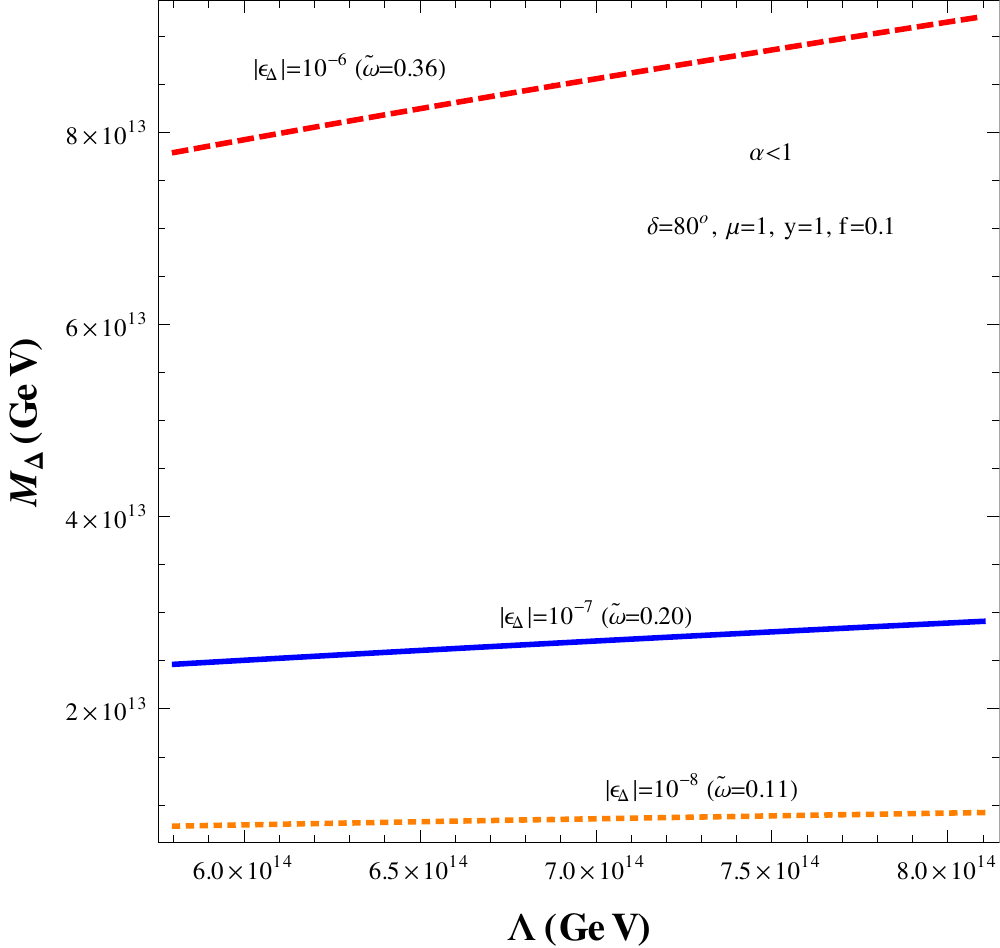}
\includegraphics[scale=0.7]{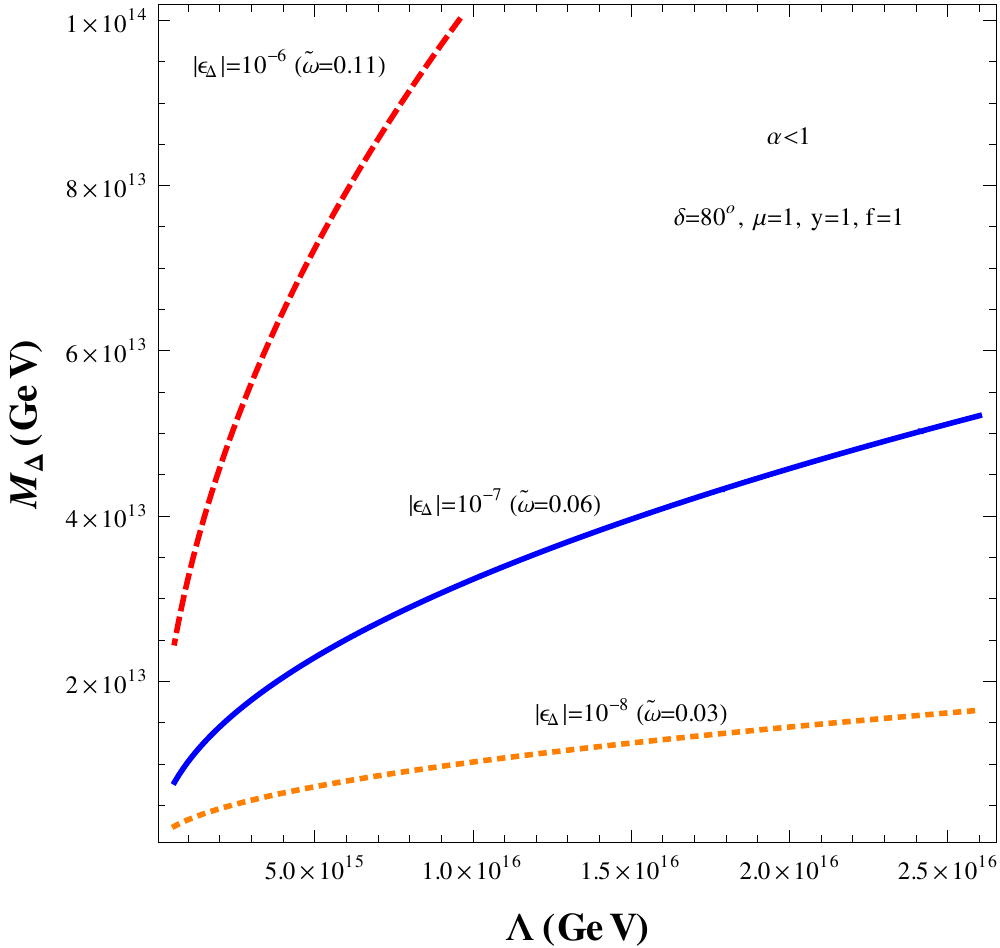}
\caption{{\small Contours of different values of $ \epsilon_{\mb}$ in
                the $M_{\mb} - \Lambda $ plane for $\alpha < 1$.
         }}
\label{c1}
\end{center}
\end{figure}

We further define $v_{\sm}/\Lambda=f\tilde{\omega}$ where $f$ serves as a 
relative measure of the vevs. With this, the expression of $\epsilon_{\mb}$ takes the form
\begin{eqnarray}\label{ep2}
 \epsilon_{\mb} &=& -\frac{\alpha^2}{8\pi v^2(1-\alpha^2)}
 \frac{
 k f (x_1-x'_1)\sin\alpha_{\sm}(\mu \Lambda/M^2_{\mb})
 }
 {
 \left[(3 f^2/ M^2_{\mb}) \left(x_1^2+x_1'^2+2x_1x_1'\cos2\alpha_{\sm}\right)
   +(\mu \Lambda/M^2_{\mb})^2\right]
 },
\end{eqnarray}
which is $\tilde{\omega}$ independent.
The expression for $|d|$ as obtained from Eq. \ref{modd} can be written as 
\begin{equation}\label{dsq}
 |d|=2 f v ^2\tilde{\omega}^4 \frac{\mu\Lambda}{M^2_{\mb}}
        (x_1+x'_1)\cos\alpha_{\sm}\sec\phi_d.
\end{equation}
Using Eq.(\ref{ep2}), we obtain the contour plot for $\epsilon_{\mb}
=10^{-6}, 10^{-7}, 10^{-8}$ with $\mu=1, f=0.1$, $x_1=0.5$ and $x'_1=1$ which are 
shown in Fig. \ref{c1}, left panel. The electroweak vev is also inserted in the
expression. In obtaining the plots we varied $\Lambda$ above the masses of RH 
neutrinos (see Eq.(\ref{MR})). The variation of $M_{\mb}$ is also restricted 
from above by the condition that we work in regime (I) where $M_{\mb}<M_{\R_ {i=1,2,3}}$.
Fig.\ref{c1} is produced for a specific choice of $\delta=80^{\circ}(100^{\circ},260^{\circ},
280^{\circ})$ which corresponds to the solution ($\alpha=0.16, \beta=0.11$). The 
values of $\left| d \right |$ and $k$ corresponding to this set of ($\alpha,\beta$)
are found to be 0.0068 eV and 0.06 eV respectively. We have chosen $y=1$ for the left
panel of Fig.\ref{c1}. In order to keep $M_{\mb}<M_{\R i}<\Lambda$, we find 
$M_{\mb}\simeq 10^{13-14}$ GeV which produce the required amount of CP-asymmetry
which in tern can generate enough lepton asymmetry. 
 
\begin{figure}[!h]
\begin{center}
\includegraphics[scale=0.7]{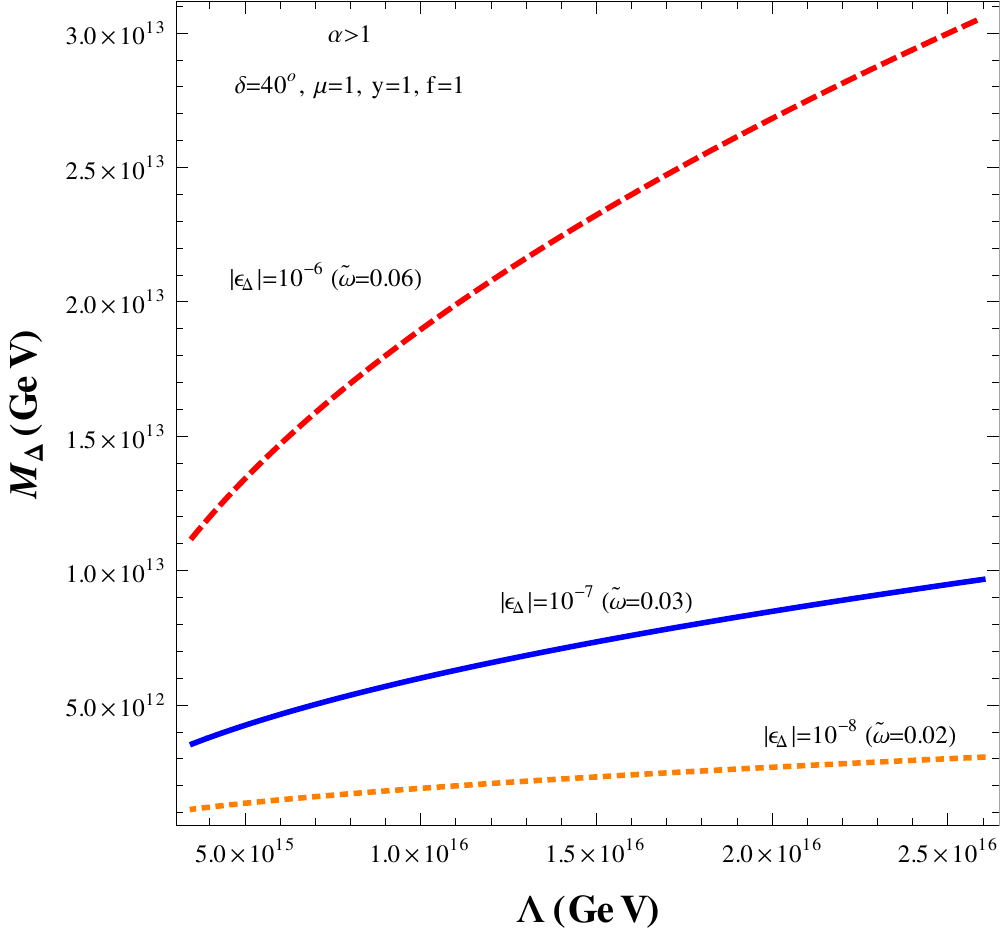}
\includegraphics[scale=0.7]{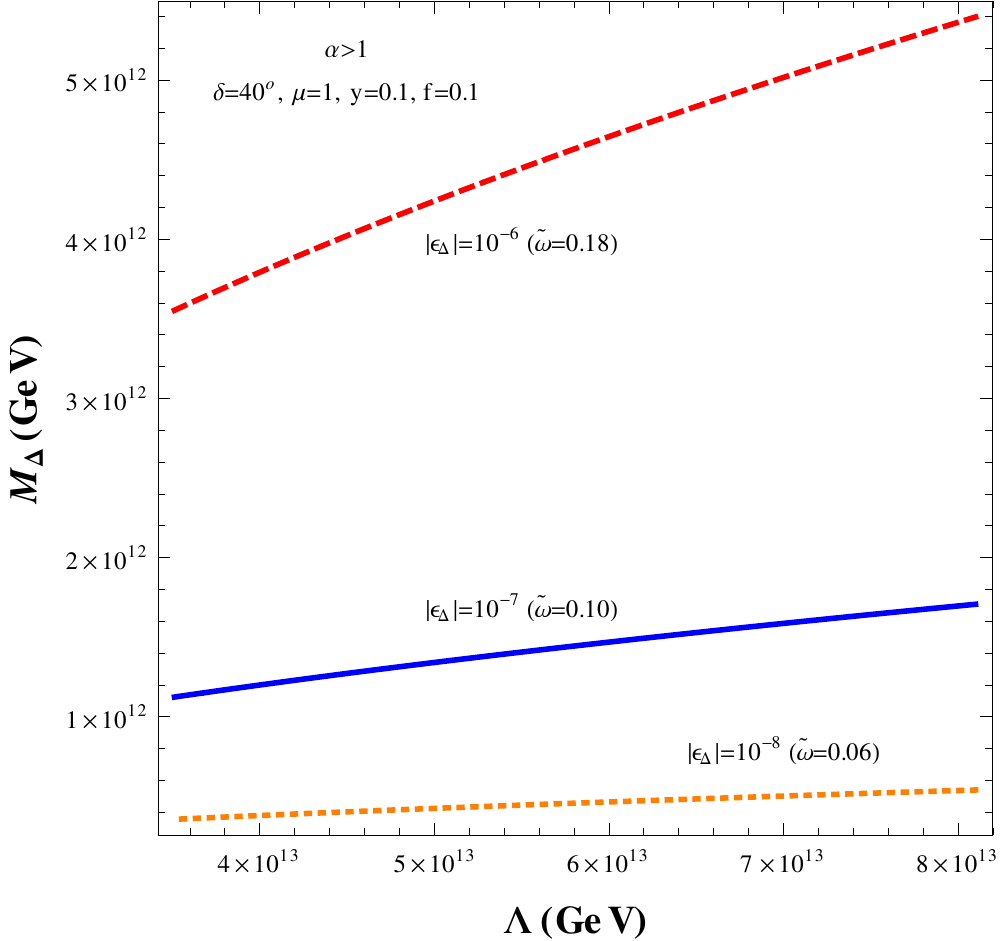}
\caption{{\small Contours of different values of $ \epsilon_{\mb}$ in the 
                     $M_{\mb} - \Lambda $ plane for $\alpha > 1$.
         }}
\label{c2}
\end{center}
\end{figure}

Note that value of $\tilde{\omega}$ can be concluded from the expression of
$|d|$ in Eq.(\ref{dsq}), for a choice of $\Lambda/M^2_{\mb}$ which produces
a $\epsilon_{\mb}$ contour. 
This is because corresponding to a specific choice of $\delta$ value,
$\left| d\right|$ is uniquely determined for the solution point $(\alpha, \beta)$.
Hence with fixed values of $x_1, x'_1, f, \mu$ (with the same values to have the 
$\epsilon_{\mb}$  contour), $\tilde{\omega}$ can be evaluated from $\left| d\right|$
for a chosen $\Lambda/M^2_{\mb}$. It turns out that $\tilde{\omega}$ has a unique 
value for a specific $\epsilon_{\mb}$ for both the panels of Fig.\ref{c1}. For 
example, with $\epsilon_{\mb}=10^{-7}$, we need $\tilde{\omega}=0.2$, while to 
have $\epsilon_{\mb}=10^{-6}$, $\tilde{\omega}$ required to be 0.36. These
$\tilde{\omega}$ values are  provided in first bracket in each figure beside
the $\epsilon_{\mb}$ value. The reason is the following. For the specified range of
$\Lambda$ ($i.e. ~M_{\mb}<M_{\R i}<\Lambda$), it follows that the first bracketed term
in the denominator of Eq.(\ref{ep2}) is almost negligible compared to the second 
term (with the choice of $x_1, x'_1, f, \mu$ as mentioned before) and hence effectively 
 \begin{equation}\label{eap}
 \epsilon_{\mb}\simeq -\frac{\alpha^2}{8\pi v^2(1-\alpha^2)}
 k f (x_1-x'_1)\sin\alpha_{\sm} \frac{M^2_{\mb}}{{\mu \Lambda}}.
  \end{equation}
 Therefore for a typical choice of $\epsilon_{\mb}$, $\Lambda/M^2_{\mb}$ is almost fixed
 and then $|d|$ expression in Eq.(\ref{eap}) tells that $\tilde{\omega}$ also is almost 
 fixed. In the right panel of Fig.\ref{c1}, we take $y=1,f=1$ and draw the contours for
 $\epsilon_{\mb}$ while $x_1, x'_1, \mu$ are fixed at their previous values considered
 for generating plots in the left panel. In this case, $M_{\mb}$ turns out to be 
 $10^{13-14}$ GeV.
\begin{figure}[!h]
\begin{center}
\includegraphics[scale=0.7]{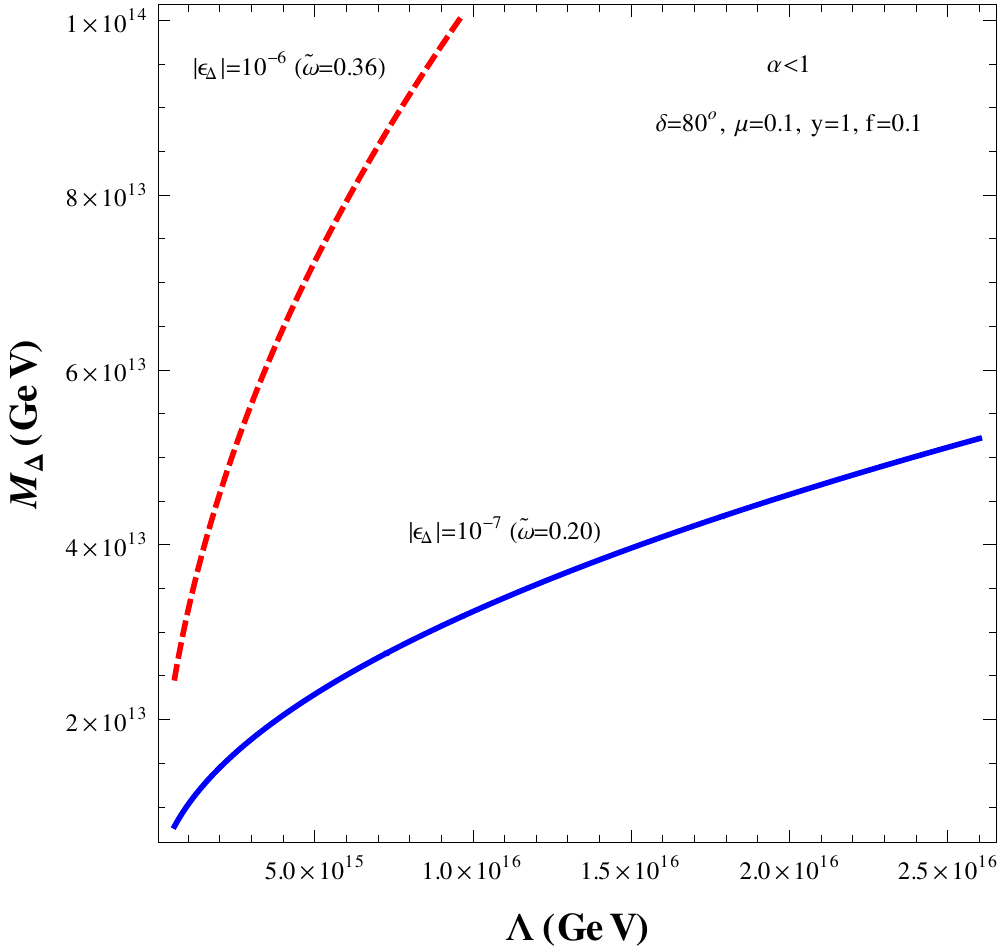}
\includegraphics[scale=0.7]{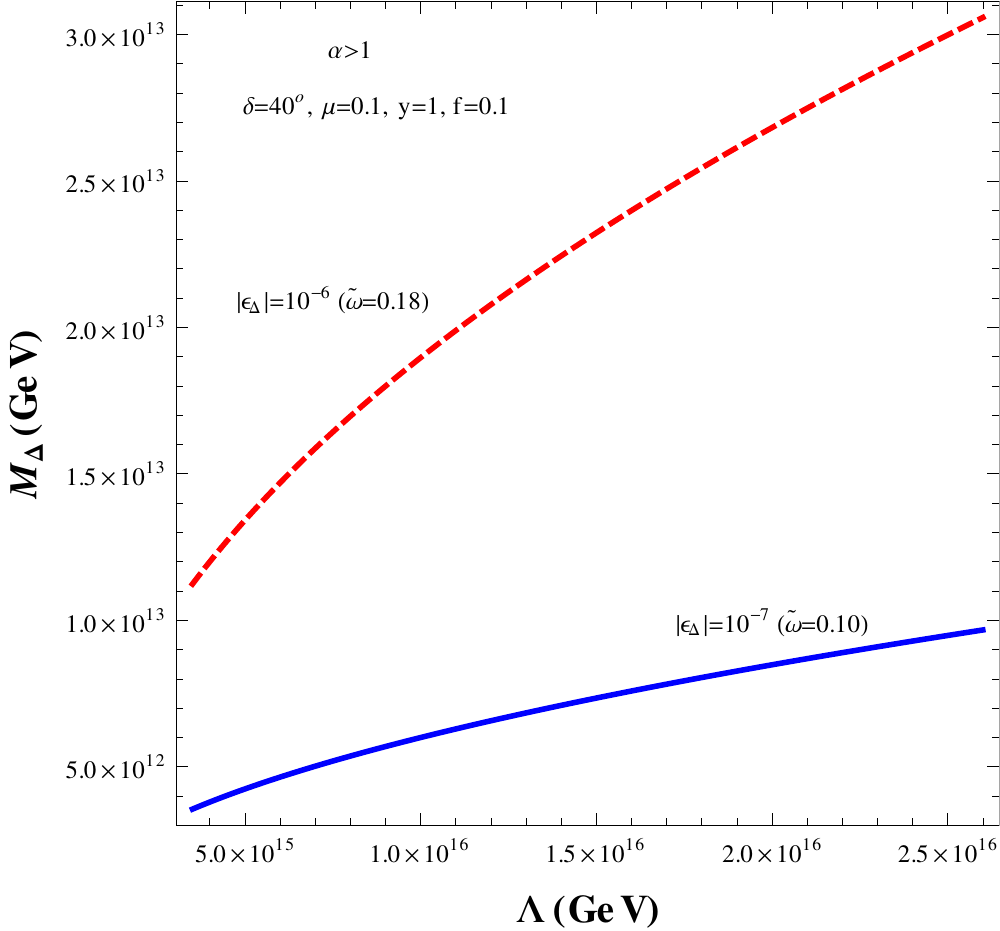}
\caption{{\small Contours of different values of $ \epsilon_{\mb}$ in the 
                     $M_{\mb} - \Lambda $ plane for $\alpha<1$ (left panel) and 
                   $\alpha>1$ (right panel) with small $\mu (=0.1)$.
         }}
\label{smu12}
\end{center}
\end{figure}

Similarly contours for $\epsilon_{\mb}$ are drawn in Fig.\ref{c2} for $\alpha>1$ case.
Correspondingly we have used solutions
of ($\alpha = 1.36,\beta = 0.53$) and the value of $k = 0.02$ eV and $|d| = 0.01$ eV 
are taken for $\delta=40^{\circ}$ (also for $140^{\circ}, 220^{\circ},320^{\circ}$).
We obtain somewhat lighter $M_{\mb}$ as correspond to the case with $\alpha<1$. 
In Fig.\ref{smu12} similar contour plots for $\epsilon_{\mb}$ are exercised with $\mu$
at some lower values, fixed at $\mu=0.1$ along with $f=0.1$ for both $\alpha<1$ and
$\alpha>1$. 

\begin{figure}[!h]
\begin{center}
\includegraphics[scale=0.7]{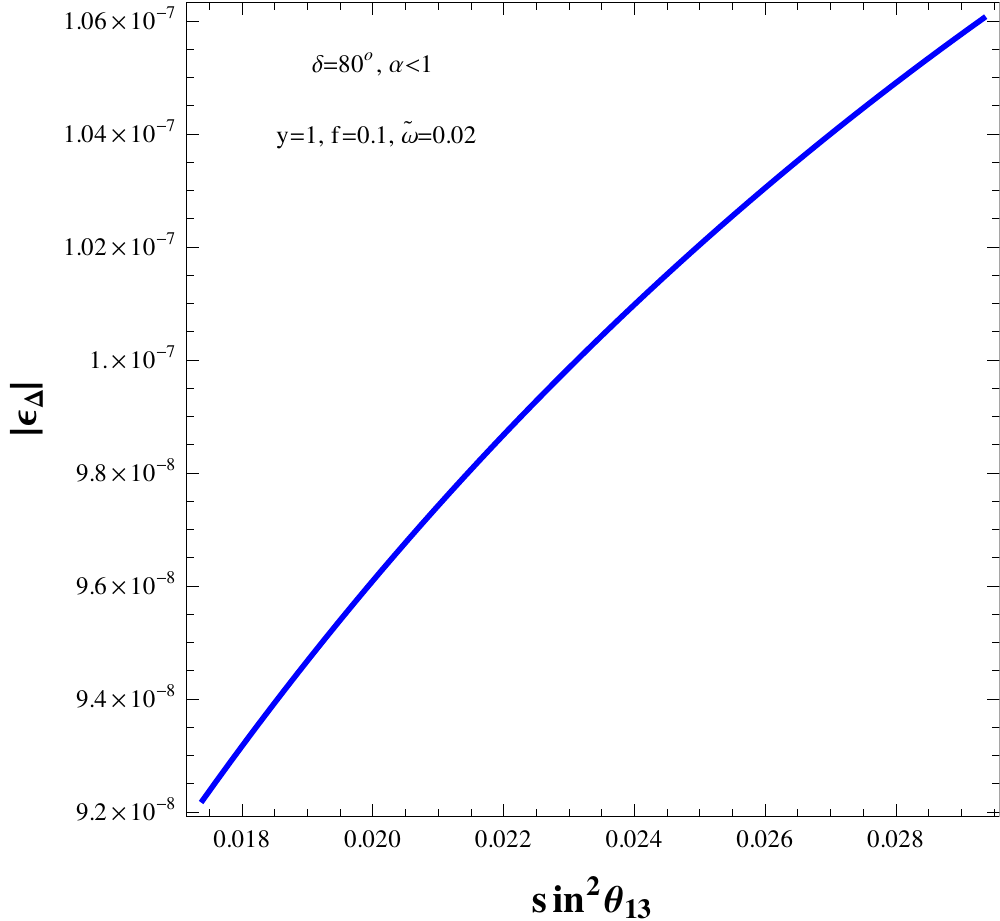}
\includegraphics[scale=0.7]{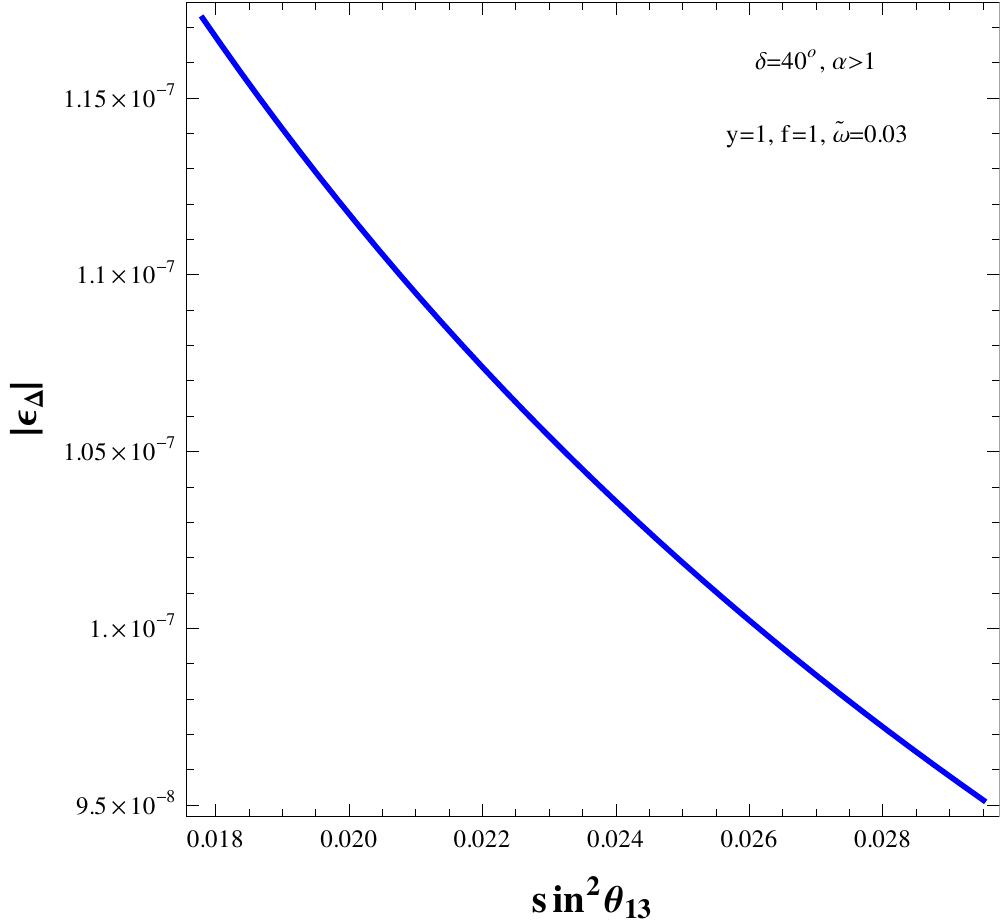}
\caption{{\small $\epsilon_{\mb}$ vs $\sin^2\theta_{13}$ for $\alpha<1$ (left panel) and 
 $\alpha>1$ (right panel).
         }}
\label{ccp}
\end{center}
\end{figure}

So overall we have found that enough $\epsilon_{\mb}$ can be created so as to achieve
the required lepton asymmetry through $\frac{n_{\Ll}}{n_{\gamma}}=\epsilon_{\mb}
\frac{n_{\mb}}{n_{\gamma}}D$ with $n_{\mb}=n_{\mb_0}+n_{\mb_+}+n_{\mb_{++}}$ is 
the total number density of the triplet and $D$ is the efficiency factor. 
After converting it into baryon asymmetry by the sphaleron process, ${n_{\B}}/{n_{\gamma}}$
is given by $\frac{n_{\B}}{n_{\gamma}}\simeq -0.03\epsilon_{\mb}D$. $D$ depends 
on the satisfaction of the out-of-equilibrium condition ($\Gamma_{\mb}\leq H|_{T=M_{\mb}}$).
Being $SU(2)_{L}$ triplet, it also contains the gauge interactions. Hence the scattering
like $\Delta\Delta\rightarrow$ SM particles can be crucial \cite{Fry:1980bc,Ma:1998sq}. 
In \cite{Hambye:2000ui,Hambye:2003ka,Hambye:2004fn,Hambye:2003rt,Hambye:2005tk}, it has
been argued that even if the triplet mass ($M_{\mb}$) is much below $10^{14}$ GeV, the 
triplet leptogenesis mechanism considered here is not affected much by the gauge mediated
scatterings. However the exact estimate of $D$ requires to solve the Boltzmann equations
in detail which is beyond the scope of the present work. However analysis toward 
evaluating $D$ in this sort of framework (where a single triplet is present and RH 
neutrinos are in the loop for generating $\epsilon_{\mb}$) exits in \cite{Hambye:2005tk}. 
Following \cite{Hambye:2005tk}, we note that with the effective type-II mass 
$\tilde{m}_{\mb}\left(\equiv\sqrt{{\rm Tr}(m_{{\nu}}^{\iii\dagger}m_{\nu}^{\iii})}\right)
\sim (0.01-0.02)$ eV, the efficiency $D$ is of the order of $10^{-3}$. In 
estimating\footnote{It is possible to recast Eq.(\ref{ep0}) as 
$\epsilon_{\mb}=-\frac{1}{8\pi}\frac{M_{\mb}}{v^2}\sqrt{B_L B_H}\frac{{\rm Tr}
(m_{{\nu}}^{\ii\dagger}m_{\nu}^{\iii})}{\tilde{m}_{\mb}}$ with the consideration 
$M_{\mb}<M_{\R k}$. Here $B_{L}$ and $B_{H}$ are corresponding branching ratio's of 
decay of the triplet into two leptons and two scalar doublets. 
} $\tilde{m}_{\mb}$, we have considered all the parameters in a range (mentioned within
Fig. \ref{c1}-\ref{c2}) so as to produce $\epsilon_{\mb}$ of order $10^{-6}$ as shown in Fig. \ref{c1}-\ref{c2}.

Now, using the approximated expression as given by Eq.(\ref{eap}) we can obtain
variation of $\epsilon_{\mb}$ against $\sin^2\theta_{13}$ as given in Fig. \ref{ccp}.
In doing so we have substituted $\mu \Lambda/M^2_{\mb}$  from Eq.(\ref{dsq}) in
Eq.(\ref{eap}). Then as discussed in the previous section, using solutions for 
$\alpha,\beta$ for 3$\sigma$ range of $\sin^2\theta_{13}$ for fixed $\delta$
we have obtained Fig.\ref{ccp} for both $\alpha<1$ and $\alpha>1$. Here
the left panel is for $\delta=80^{\circ}(100^{\circ}, 260^{\circ},280^{\circ})$
and right panel in for  $\delta=40^{\circ}(140^{\circ},220^{\circ},320^{\circ})$.

\hspace{.3cm}
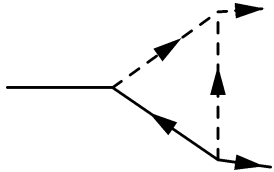
\begin{figure}[!h]
\begin{center}
\begin{fmffile}{n5}
\begin{fmfgraph*}(110,60)
\fmfleft{i1}
\fmfright{o1,o2}
\fmf{plain}{i1,v1}
\fmf{fermion,tension=.5}{v2,v1}
\fmf{scalar,tension=.5}{v1,v3}
\fmf{scalar,tension=0,label=$\Delta$}{v2,v3}
\fmf{fermion}{v2,o1}
\fmf{scalar}{v3,o2}
\fmflabel{$N_{\R_k}$}{i1}
\fmflabel{$L_i$}{o1}
\fmflabel{$H^{\ast}$}{o2}
\fmffreeze
\fmfshift{60down}{v2}
\fmfshift{71up}{v3}
\end{fmfgraph*}
\end{fmffile}
\vspace{.3cm}
\caption{ One-loop diagram for decay of RH neutrinos }\label{figl2}
\end{center}
\end{figure}


We now discuss the option II, when RH neutrinos are lighter than $M_{\mb}$. The contribution toward the 
CP-asymmetry parameter generated from the decay of the lightest  neutrino is given by 
\begin{eqnarray}
\epsilon_{\N_1}&=&-\frac{1}{8 \pi v^2} M_{\R_{1}}
                       \frac{\sum_{il}{\rm Im}[(\hat{Y}_{\D})_{1i} 
                             (\hat{Y}_{\D})_{1l}(m^{II*}_\nu)_{il}]}
                            {\sum_i | (\hat{Y}_{\D})_{1i}|^2} ,    \\
            &=&-\frac{M_{\R_1}}{2}\frac{1}{8\pi v^2}|d|\sin\phi_d, \\
\epsilon_{\N_2}&=&M_{\R_2}\frac{1}{8\pi v^2}|d|\sin\phi_d \hspace{0.5cm} {\rm and} \hspace{.5cm}
\epsilon_{\N_3}= \pm \frac{M_{\R_3}}{2}\frac{1}{8\pi v^2}|d|\sin\phi_d   .
\end{eqnarray}
where we have used $m_{\nu}^{\iii}$ from Eq.(\ref{d}). In the above, $`+$' and $`-$' sign stands for $\alpha>1$ and 
$\alpha<1$ cases respectively in computation of $\epsilon_{\N_3}$. Note that in the present scenario
the RH neutrino masses are not entirely hierarchical, rather they are closely placed. therefore the 
total baryon asymmetry from the decay of the three RH neutrinos is to be estimated as 
$\left|\frac{n_{\B}}{s}\right|=1.48\times10^{-3}\sum_i\epsilon_{\N_i}D_{\N_i}$, where 
$D_{\N_i}$ is the respective efficiency factor. It turns out that with the same $D_{\N_i}$ 
for $i=1,2,3$, $\sum_i\epsilon_{\N_i}=0$ as a result (using $M_{\R_i}$ from Eq.(\ref{MR}))
of the specific flavor structure considered. Therefore it is expected that the lepton asymmetry
would be suppressed in this case. Also in this case $M_{\R_i}<M_{\mb}$, which can be 
obtained by considering smaller value of the Yukawa coupling $y$ (as to generate the required
$|d|$, specific values of $\alpha,\beta,k$ are already chosen ). This could reduce the individual
$\epsilon_{\N_i}$. We conclude this contribution ($\epsilon_{\N}$) as a subdominant to $\epsilon_{\mb}$.



\section{Conclusion}\label{conlc}
We have considered a flavor symmetric framework for generating light neutrino masses 
and mixing through type-II seesaw mechanism. In realizing it, we have introduced three
SM singlet RH neutrinos, one $SU(2)_L$ triplet and few flavon fields. The RH neutrinos 
contribute to the type-I term, which guided by the $A_4 \times Z_4 \times Z_3$ symmetry
of the model produces a TBM mixing pattern. Then we have shown that the typical flavor
structure resulted from the model can generate nonzero $\theta_{13}$. In this framework,
all the couplings are considered to be real. The CP symmetry is violated spontaneously
by the complex vev of a single SM singlet field, while other flavons have real vevs. 
Interestingly this particular field is involved only in the pure type-II term. Hence
the triplet contribution not only generates the $\theta_{13}$, it is also responsible
for providing Dirac CP violating phase $\delta$. Therefore the model has the potential
to predict $\delta$ in terms of the parameters involved in neutrino masses and mixing. 
We have therefore studied the parameter space of the set-up considering that the triplet
contribution is subdominant or at most comparable to the type-I term. The model indicates 
the values of $\delta$ to be in the range $72^{\circ}-82^{\circ}$,  $98^{\circ}-108^{\circ}$,
$252^{\circ}-262^{\circ}$, $278^{\circ}-288^{\circ}$ for $\alpha<1$ and $\delta=0^{\circ}-63^{\circ}$, 
$117^{\circ}-180^{\circ}$, $180^{\circ}-243^{\circ}$, $297^{\circ}-360^{\circ}$ for $\alpha>1$. 
However  $\delta=0$ (and hence $\pi,2\pi$) is disfavored in our scenario as in that case no CP
violation would be present. Also  $\delta=\pi/2, 3\pi/2$ are excluded here. These ranges can be
tested in future neutrino experiments. We provide an estimate for the $J_{CP}$. The sum of the
neutrino masses are also evaluated. It turns out that the scenario works with normal hierarchical
masses of light neutrinos. 
We have also studied leptogenesis in this model. As the type-I contribution to the light neutrino
mass does not involve any CP violating phase, RH neutrinos decay can not contribute to the lepton
asymmetry in the conventional way. We have found the triplet decay with the virtual RH neutrino in
the loop can produce enough lepton asymmetry. 



\begin{thebibliography}{00}
\bibitem{Minkowski:1977sc} 
  P.~Minkowski,
  Phys.\ Lett.\ B {\bf 67}, 421 (1977).

\bibitem{GellMann:1980vs} 
  M.~Gell-Mann, P.~Ramond and R.~Slansky,
  Conf.\ Proc.\ C {\bf 790927}, 315 (1979)
  [arXiv:1306.4669 [hep-th]].
  
\bibitem{Mohapatra:1979ia} 
  R.~N.~Mohapatra and G.~Senjanovic,
  Phys.\ Rev.\ Lett.\  {\bf 44}, 912 (1980).

\bibitem{Yanagida:1980xy} 
  T.~Yanagida,
  Prog.\ Theor.\ Phys.\  {\bf 64}, 1103 (1980).
%
\bibitem{Harrison:1999cf} 
  P.~F.~Harrison, D.~H.~Perkins and W.~G.~Scott,
  Phys.\ Lett.\ B {\bf 458}, 79 (1999)
  [hep-ph/9904297].
\bibitem{Ma:2004zv} 
  E.~Ma,
  Phys.\ Rev.\ D {\bf 70}, 031901 (2004)
  [hep-ph/0404199].

\bibitem{Altarelli:2005yp} 
  G.~Altarelli and F.~Feruglio,
  Nucl.\ Phys.\ B {\bf 720}, 64 (2005)
  [hep-ph/0504165].

\bibitem{Altarelli:2005yx} 
  G.~Altarelli and F.~Feruglio,
  Nucl.\ Phys.\ B {\bf 741}, 215 (2006)
  [hep-ph/0512103].
\bibitem{Abe:2011fz} 
  Y.~Abe {\it et al.}  [DOUBLE-CHOOZ Collaboration],
  Phys.\ Rev.\ Lett.\  {\bf 108}, 131801 (2012)
  [arXiv:1112.6353 [hep-ex]].

\bibitem{DayaBay} 
  F.~P.~An {\it et al.}  [DAYA-BAY Collaboration],
  Phys.\ Rev.\ Lett.\  {\bf 108}, 171803 (2012)
  [arXiv:1203.1669 [hep-ex]];
  F.~P.~An {\it et al.}  [ Daya Bay Collaboration],
  arXiv:1406.6468 [hep-ex].

\bibitem{Ahn:2012nd} 
  J.~K.~Ahn {\it et al.}  [RENO Collaboration],
  Phys.\ Rev.\ Lett.\  {\bf 108}, 191802 (2012)
  [arXiv:1204.0626 [hep-ex]].


\bibitem{Abe:2013hdq} 
  K.~Abe {\it et al.}  [T2K Collaboration],
  Phys.\ Rev.\ Lett.\  {\bf 112}, 061802 (2014)
  [arXiv:1311.4750 [hep-ex]].
  
\bibitem{Karmakar:2014dva} 
  B.~Karmakar and A.~Sil,
  Phys.\ Rev.\ D {\bf 91}, 013004 (2015)
  [arXiv:1407.5826 [hep-ph]].
  
\bibitem{Joshipura:1999is} 
  A.~S.~Joshipura and E.~A.~Paschos,
  hep-ph/9906498.

\bibitem{Joshipura:2001ya} 
  A.~S.~Joshipura, E.~A.~Paschos and W.~Rodejohann,
  Nucl.\ Phys.\ B {\bf 611}, 227 (2001)
  [hep-ph/0104228].
  
\bibitem{Bajc:2002iw} 
  B.~Bajc, G.~Senjanovic and F.~Vissani,
  Phys.\ Rev.\ Lett.\  {\bf 90}, 051802 (2003)
  [hep-ph/0210207].
  
\bibitem{Antusch:2004xd} 
  S.~Antusch and S.~F.~King,
  Nucl.\ Phys.\ B {\bf 705}, 239 (2005)
  [hep-ph/0402121].
  
\bibitem{Antusch:2004xy} 
  S.~Antusch and S.~F.~King,
  Phys.\ Lett.\ B {\bf 597}, 199 (2004)
  [hep-ph/0405093].
  
\bibitem{Sahu:2004ny} 
  N.~Sahu and S.~U.~Sankar,
  Phys.\ Rev.\ D {\bf 71}, 013006 (2005)
  [hep-ph/0406065].
    
  
\bibitem{Rodejohann:2004qh} 
  W.~Rodejohann and Z.~z.~Xing,
  Phys.\ Lett.\ B {\bf 601}, 176 (2004)
  [hep-ph/0408195].
  

\bibitem{Chen:2005jm} 
  S.~L.~Chen, M.~Frigerio and E.~Ma,
  Nucl.\ Phys.\ B {\bf 724}, 423 (2005)
  [hep-ph/0504181].
  
\bibitem{Bertolini:2005qb} 
  S.~Bertolini and M.~Malinsky,
  Phys.\ Rev.\ D {\bf 72}, 055021 (2005)
  [hep-ph/0504241].
  

\bibitem{Akhmedov:2006de} 
  E.~K.~Akhmedov and M.~Frigerio,
  JHEP {\bf 0701}, 043 (2007)
  [hep-ph/0609046].



\bibitem{Rodejohann:2004cg} 
  W.~Rodejohann,
  Phys.\ Rev.\ D {\bf 70}, 073010 (2004)
  [hep-ph/0403236].
  
\bibitem{Gu:2006wj} 
  P.~H.~Gu, H.~Zhang and S.~Zhou,
  Phys.\ Rev.\ D {\bf 74}, 076002 (2006)
  [hep-ph/0606302].
  
\bibitem{Abada:2008gs} 
  A.~Abada, P.~Hosteins, F.~X.~Josse-Michaux and S.~Lavignac,
  Nucl.\ Phys.\ B {\bf 809}, 183 (2009)
  [arXiv:0808.2058 [hep-ph]].

  
  

  
\bibitem{Borah:2013bza} 
  D.~Borah and M.~K.~Das,
  Phys.\ Rev.\ D {\bf 90}, no. 1, 015006 (2014)
  [arXiv:1303.1758 [hep-ph]].
  
\bibitem{Borah:2014fga} 
  D.~Borah,
  Int.\ J.\ Mod.\ Phys.\ A {\bf 29}, 1450108 (2014)
  [arXiv:1403.7636 [hep-ph]].
  
\bibitem{Borah:2014bda} 
  M.~Borah, D.~Borah, M.~K.~Das and S.~Patra,
  Phys.\ Rev.\ D {\bf 90}, no. 9, 095020 (2014)
  [arXiv:1408.3191 [hep-ph]].
    
\bibitem{Kalita:2014vxa} 
  R.~Kalita, D.~Borah and M.~K.~Das,
  Nucl.\ Phys.\ B {\bf 894}, 307 (2015)
  [arXiv:1412.8333 [hep-ph]]. 
  
\bibitem{Pramanick:2015qga} 
  S.~Pramanick and A.~Raychaudhuri,
  arXiv:1508.02330 [hep-ph].
\bibitem{Magg:1980ut} 
  M.~Magg and C.~Wetterich,
  Phys.\ Lett.\ B {\bf 94}, 61 (1980).
  
\bibitem{Lazarides:1980nt} 
  G.~Lazarides, Q.~Shafi and C.~Wetterich,
  Nucl.\ Phys.\ B {\bf 181}, 287 (1981).
  
\bibitem{Mohapatra:1980yp} 
  R.~N.~Mohapatra and G.~Senjanovic,
  Phys.\ Rev.\ D {\bf 23}, 165 (1981).
  
 \bibitem{Schechter:1980gr} 
   J.~Schechter and J.~W.~F.~Valle,
   Phys.\ Rev.\ D {\bf 22}, 2227 (1980).
\bibitem{Capozzi:2013csa} 
  F.~Capozzi, G.~L.~Fogli, E.~Lisi, A.~Marrone, D.~Montanino and A.~Palazzo,
  Phys.\ Rev.\ D {\bf 89}, no. 9, 093018 (2014)
  [arXiv:1312.2878 [hep-ph]].

\bibitem{Gonzalez-Garcia:2014bfa} 
  M.~C.~Gonzalez-Garcia, M.~Maltoni and T.~Schwetz,
  JHEP {\bf 1411}, 052 (2014)
  [arXiv:1409.5439 [hep-ph]].
    
\bibitem{Forero:2014bxa} 
  D.~V.~Forero, M.~Tortola and J.~W.~F.~Valle,
  Phys.\ Rev.\ D {\bf 90}, no. 9, 093006 (2014)
  [arXiv:1405.7540 [hep-ph]].
  
\bibitem{Lee:1973iz} 
  T.~D.~Lee,
  Phys.\ Rev.\ D {\bf 8}, 1226 (1973).
  
\bibitem{Nelson:1983zb} 
  A.~E.~Nelson,
  Phys.\ Lett.\ B {\bf 136}, 387 (1984).

\bibitem{Barr:1984qx} 
  S.~M.~Barr,
  Phys.\ Rev.\ Lett.\  {\bf 53}, 329 (1984).
  
\bibitem{Harvey:1980je} 
  J.~A.~Harvey, P.~Ramond and D.~B.~Reiss,
  Phys.\ Lett.\ B {\bf 92}, 309 (1980).
  
  
\bibitem{Harvey:1981hk} 
  J.~A.~Harvey, D.~B.~Reiss and P.~Ramond,
  Nucl.\ Phys.\ B {\bf 199}, 223 (1982).
  
\bibitem{Branco:1980sz} 
  G.~C.~Branco,
  Phys.\ Rev.\ D {\bf 22}, 2901 (1980).
  
\bibitem{Bento:1990wv} 
  L.~Bento and G.~C.~Branco,
  Phys.\ Lett.\ B {\bf 245}, 599 (1990).
  
\bibitem{Bento:1991ez} 
  L.~Bento, G.~C.~Branco and P.~A.~Parada,
  Phys.\ Lett.\ B {\bf 267}, 95 (1991).
\bibitem{Branco:2003rt} 
  G.~C.~Branco, P.~A.~Parada and M.~N.~Rebelo,
  hep-ph/0307119.
  
\bibitem{Branco:2012vs} 
  G.~C.~Branco, R.~Gonzalez Felipe, F.~R.~Joaquim and H.~Serodio,
  Phys.\ Rev.\ D {\bf 86}, 076008 (2012)
  [arXiv:1203.2646 [hep-ph]].

  




\bibitem{Branco:2001pq} 
  G.~C.~Branco, T.~Morozumi, B.~M.~Nobre and M.~N.~Rebelo,
  Nucl.\ Phys.\ B {\bf 617}, 475 (2001)
  [hep-ph/0107164].
 
\bibitem{Araki:2012hb} 
  T.~Araki and H.~Ishida,
  PTEP {\bf 2014}, no. 1, 013B01 (2014)
  [arXiv:1211.4452 [hep-ph]].
 

\bibitem{Ahn:2013mva} 
  Y.~H.~Ahn, S.~K.~Kang and C.~S.~Kim,
  Phys.\ Rev.\ D {\bf 87}, no. 11, 113012 (2013)
  [arXiv:1304.0921 [hep-ph]].
  
\bibitem{Kim:2015etv} 
  J.~E.~Kim and S.~Nam,
  arXiv:1506.08491 [hep-ph].
  
  
%


   

  
  
  
  
  
\bibitem{Achiman:2004qf} 
  Y.~Achiman,
  Phys.\ Lett.\ B {\bf 599}, 75 (2004)
  [hep-ph/0403309].
  
\bibitem{Achiman:2007qz} 
  Y.~Achiman,
  Phys.\ Lett.\ B {\bf 653}, 325 (2007)
  [hep-ph/0703215].
  
\bibitem{Frank:2004xt} 
  M.~Frank,
  Phys.\ Rev.\ D {\bf 70}, 036004 (2004).

  
  

\bibitem{Chen:2004ww} 
  M.~C.~Chen and K.~T.~Mahanthappa,
  Phys.\ Rev.\ D {\bf 71}, 035001 (2005)
  [hep-ph/0411158].


\bibitem{Sahu:2005qm} 
  N.~Sahu and S.~U.~Sankar,
  Nucl.\ Phys.\ B {\bf 724}, 329 (2005)
  [hep-ph/0501069].


\bibitem{Chao:2007rm} 
  W.~Chao, S.~Luo and Z.~z.~Xing,
  Phys.\ Lett.\ B {\bf 659}, 281 (2008)
  [arXiv:0704.3838 [hep-ph]].

































  
\bibitem{Agashe:2014kda} 
  K.~A.~Olive {\it et al.}  [Particle Data Group Collaboration],
  Chin.\ Phys.\ C {\bf 38}, 090001 (2014).
  
  
  
  
  
  
  
  
\bibitem{NuExpt} 
  S.~Fukuda {\it et al.}  [Super-Kamiokande Collaboration],
  Phys.\ Lett.\ B {\bf 539}, 179 (2002)
  [hep-ex/0205075];
  Y.~Ashie {\it et al.}  [Super-Kamiokande Collaboration],
  Phys.\ Rev.\ D {\bf 71}, 112005 (2005)
  [hep-ex/0501064].
  P.~Adamson {\it et al.}  [MINOS Collaboration],
  Phys.\ Rev.\ Lett.\  {\bf 106}, 181801 (2011)
  [arXiv:1103.0340 [hep-ex]].
  T.~Araki {\it et al.}  [KamLAND Collaboration],
  Phys.\ Rev.\ Lett.\  {\bf 94}, 081801 (2005)
  [hep-ex/0406035].
  
  

  







  
  
  
\bibitem{Ade:2013zuv} 
  P.~A.~R.~Ade {\it et al.}  [Planck Collaboration],
  arXiv:1303.5076 [astro-ph.CO].
  
\bibitem{Asakura:2014lma} 
  K.~Asakura {\it et al.} [KamLAND-Zen Collaboration],
  AIP Conf.\ Proc.\  {\bf 1666}, 170003 (2015)
  [arXiv:1409.0077 [physics.ins-det]].
  
\bibitem{Albert:2014awa} 
  J.~B.~Albert {\it et al.} [EXO-200 Collaboration],
  Nature {\bf 510}, 229 (2014)
  [arXiv:1402.6956 [nucl-ex]].
  
  
  
  
  
  
  
  
  
  
  
\bibitem{King:2011zj} 
  S.~F.~King and C.~Luhn,
  JHEP {\bf 1109}, 042 (2011)
  [arXiv:1107.5332 [hep-ph]].
  
  
  
  
  
  

 
\bibitem{Altarelli:2012ss} 
  G.~Altarelli, F.~Feruglio and L.~Merlo,
  Fortsch.\ Phys.\  {\bf 61}, 507 (2013)
  [arXiv:1205.5133 [hep-ph]].
 
 
 
 
 
 
 
 
 
 
 

  
  
  
\bibitem{Davidson:2008bu} 
 For a review, see 
  S.~Davidson, E.~Nardi and Y.~Nir,
  Phys.\ Rept.\  {\bf 466}, 105 (2008)
  [arXiv:0802.2962 [hep-ph]] and references there in.
  
  
  

  
  
\bibitem{O'Donnell:1993am} 
  P.~J.~O'Donnell and U.~Sarkar,
  Phys.\ Rev.\ D {\bf 49}, 2118 (1994)
  [hep-ph/9307279].
  
  

  
    
  
\bibitem{Ma:1998dx} 
  E.~Ma and U.~Sarkar,
  Phys.\ Rev.\ Lett.\  {\bf 80}, 5716 (1998)
  [hep-ph/9802445];
  
\bibitem{Hambye:2000ui} 
  T.~Hambye, E.~Ma and U.~Sarkar,
  Nucl.\ Phys.\ B {\bf 602}, 23 (2001)
  [hep-ph/0011192];
  
  
\bibitem{Hambye:2005tk} 
  T.~Hambye, M.~Raidal and A.~Strumia,
  Phys.\ Lett.\ B {\bf 632}, 667 (2006)
  [hep-ph/0510008].
  
\bibitem{Branco:2011zb} 
  G.~C.~Branco, R.~G.~Felipe and F.~R.~Joaquim,
  Rev.\ Mod.\ Phys.\  {\bf 84}, 515 (2012)
  [arXiv:1111.5332 [hep-ph]].
  
\bibitem{AristizabalSierra:2009ex} 
  D.~Aristizabal Sierra, F.~Bazzocchi, I.~de Medeiros Varzielas, L.~Merlo and S.~Morisi,
  Nucl.\ Phys.\ B {\bf 827}, 34 (2010)
  [arXiv:0908.0907 [hep-ph]].
  
\bibitem{AristizabalSierra:2012js} 
  D.~Aristizabal Sierra and I.~de Medeiros Varzielas,
  Fortsch.\ Phys.\  {\bf 61}, 645 (2013)
  [arXiv:1205.6134 [hep-ph]].

  

\bibitem{Hambye:2003ka} 
  T.~Hambye and G.~Senjanovic,
  Phys.\ Lett.\ B {\bf 582}, 73 (2004)
  [hep-ph/0307237].



  

  

  

   
\bibitem{Antusch:2007km} 
  S.~Antusch,
  Phys.\ Rev.\ D {\bf 76}, 023512 (2007)
  [arXiv:0704.1591 [hep-ph]].
  

  
  
\bibitem{Akhmedov:2008tb} 
  E.~K.~Akhmedov and W.~Rodejohann,
  JHEP {\bf 0806}, 106 (2008)
  [arXiv:0803.2417 [hep-ph]].
  
\bibitem{AristizabalSierra:2011ab} 
  D.~Aristizabal Sierra, F.~Bazzocchi and I.~de Medeiros Varzielas,
  Nucl.\ Phys.\ B {\bf 858}, 196 (2012)
  [arXiv:1112.1843 [hep-ph]].
 

\bibitem{Sierra:2014tqa} 
  D.~Aristizabal Sierra, M.~Dhen and T.~Hambye,
  JCAP {\bf 1408}, 003 (2014)
  [arXiv:1401.4347 [hep-ph]].
  
\bibitem{Hambye:2012fh} 
  T.~Hambye,
  New J.\ Phys.\  {\bf 14}, 125014 (2012)
  [arXiv:1212.2888 [hep-ph]].
  
\bibitem{Lazarides:1998iq} 
  G.~Lazarides and Q.~Shafi,
  Phys.\ Rev.\ D {\bf 58}, 071702 (1998)
  [hep-ph/9803397].

\bibitem{Fry:1980bc} 
  J.~N.~Fry, K.~A.~Olive and M.~S.~Turner,
  Phys.\ Rev.\ D {\bf 22}, 2977 (1980).
  
\bibitem{Ma:1998sq} 
  E.~Ma, S.~Sarkar and U.~Sarkar,
  Phys.\ Lett.\ B {\bf 458}, 73 (1999)
  [hep-ph/9812276].

\bibitem{Hambye:2004fn} 
  T.~Hambye,
  hep-ph/0412053.
  
\bibitem{Hambye:2003rt} 
  T.~Hambye, Y.~Lin, A.~Notari, M.~Papucci and A.~Strumia,
  Nucl.\ Phys.\ B {\bf 695}, 169 (2004)
  [hep-ph/0312203].
  
\end{thebibliography}
\end{document}